\RequirePackage[2020-02-02]{latexrelease}
  \documentclass[amsmath,12pt,amssymb,preprint,prd,aps,nofootinbib]{revtex4}
  \usepackage{amsfonts}
  \usepackage{rotating} 
  \usepackage{graphicx} 
  \usepackage{epsfig}
  \usepackage{multirow}
  \usepackage{bm}
  \usepackage{hyperref}
  \usepackage{cleveref}
  \usepackage[toc,page]{appendix}

  \newcommand{\diracslash}[1]{#1\llap{/\kern2pt}}
  
  \newcommand{\be}{\begin{equation}}
  \newcommand{\ee}{\end{equation}}
  \newcommand{\bea}{\begin{eqnarray}}
  \newcommand{\eea}{\end{eqnarray}}
  \newcommand{\ba}[1]{\begin{array}{#1}}
  \newcommand{\ea}{\end{array}}
  
  \newcommand{\bt}{\begin{tabular}}
  \newcommand{\et}{\end{tabular}}

  \newcommand{\beas}{\begin{eqnarray*}}
  \newcommand{\eeas}{\end{eqnarray*}}

\begin{document}
\title{Magnetic moments of decuplet baryons in asymmetric strange hadronic matter at finite temperature} 

\author{\large Arvind Kumar}
\email{kumara@nitj.ac.in}
\author{\large Suneel Dutt}
\email{dutts@nitj.ac.in}
\author{\large Harleen Dahiya}
\email{dahiyah@nitj.ac.in}
 
\affiliation{Department of Physics, Dr. B R Ambedkar National Institute of Technology Jalandhar, 
 Jalandhar -- 144008, Punjab, India}

\def\be{\begin{equation}}
\def\ee{\end{equation}}
\def\bearr{\begin{eqnarray}}
\def\eearr{\end{eqnarray}}
\def\zbf#1{{\bf {#1}}}
\def\bfm#1{\mbox{\boldmath $#1$}}
\def\hf{\frac{1}{2}}
\def\kp{\zbf k+\frac{\zbf q}{2}}
\def\km{-\zbf k+\frac{\zbf q}{2}}
\def\hwo{\hat\omega_1}
\def\hwt{\hat\omega_2}

        \begin{abstract}
In the present work we study the   masses and magnetic moments of decuplet baryons in isospin asymmetric strange hadronic medium at finite temperature using chiral SU(3) quark mean field model. In the strange isospin asymmetric medium, the properties of baryons in chiral SU(3) mean field model are modified through the exchange of  scalar fields $\sigma$, $\zeta$ and $\delta$ and the vector fields $\omega$, $\rho$ and $\phi$. The scalar-isovector field $\delta$  and the vector-isovector field $\rho$ signifies the finite isospin asymmetry of the medium. We calculate the in-medium constituent quark masses and masses of decuplet baryons in asymmetric strange matter within the chiral SU(3) quark mean field model and use these as input in the constituent chiral quark model to calculate the in-medium magnetic moments of decuplet baryons for different values of isospin asymmetry and strangeness fraction of hot and dense medium. For calculating the magnetic moments of baryons,  contributions of valence quarks, quark sea and orbital angular momentum of quark sea are considered in the calculations.
\end{abstract}
	
	\maketitle
	
	\section{\label{intro}Introduction}
The theory of strong interactions,  quantum chromodynamics	(QCD) treats quarks and gluons as fundamental degrees of freedom. The dynamics of these degrees of freedom and hence the properties of hadrons (whose constituents are quarks and gluons) may be modified in the strongly interacting dense nuclear environment at zero and finite temperature. 
Investigating the medium modification of different properties of hadrons, over a wide range of temperature and density, is of interest for the research community of high energy physics as these studies have direct implications in  understanding the experimental observables which might be produced
in heavy ion collision experiments as well as in the astrophysical laboratories, the compact stars.

Exploring the impact  of finite isospin asymmetry and presence of hyperons in the strongly interacting medium can have direct influence on the properties of mesons and baryons. 
Many studies are available in the literature adopting different strategies, for example, chiral quark model, QCD sum rules, light cone QCD models etc.,  to calculate the electromagnetic properties such as magnetic moments and electromagnetic form factors of baryons in the free space \cite{Contreras2004,Araujo2004,Jones2000,Jun2007,lksharam2007,Sahu2002,Slaughter2011,Dahiya2003,JDey2000,Aliev2000}. 
Exploring the impact of finite density of nuclear matter under strong interactions on the electromagnetic properties has also got attention.
The European Muon Collaboration (EMC) effect involves the modification of nucleon structure inside the nuclei \cite{emc1}.
The evidence of medium modification of electromagnetic form factors
is provided by polarized ($\vec{e},e^{'}p$) scattering off $^{16}O$ and $^{4}He$ nuclei in MAMI and Jefferson lab experiments \cite{ Strauch2003,Paolone2010,Malace2011}.
 The anomalous magnetic moments of baryons play a significant role in determining the properties of medium subjected to finite external magnetic field \cite{Arghya2018,Aguirre2019,Sinha2013,Rabhi2011,Dexheimer2012,Rajesh2020,Rajesh22020,Parui2021,open_charm_mag_AM_SPM,charmdw_mag}. The occurrence of magnetic catalysis (MC) or inverse magnetic catalysis (IMC)
 in a nuclear
  medium with finite external magnetic field
   depends upon whether the magnetic moments of baryons are included in the study or not \cite{Arghya2018,Aguirre2019}. The anomalous magnetic moments of baryons are also important for investigating the properties of magnetars
 \cite{Manisha2022,thapa2022,Marquez2022}.

 The quark meson coupling (QMC) model and modified quark meson coupling models have been employed to study the magnetic moments of octet baryons in the nuclear medium \cite{ryu3_plb2009,ryu_G2010}.  The impact of medium modification of magnetic moments of baryons on the properties of neutron stars under strong magnetic field has been investigated in Ref. \cite{ryu1_prc2010}.
 The magnetic moment of neutron in dense magnetized medium are investigated using lowest order constraint
 variation (LOCV) method in \cite{Zeinab2018}.
 The magnetic moments of the octet, decuplet, low-lying
 charm, and low-lying bottom baryons 
 have been calculated in the nuclear
 medium using QMC model  \cite{Tsushimaptep2022}.
 The electromagnetic form factors of baryons has also been studied in the nuclear medium using covariant spectator quark model \cite{Ramalho2013}.
 The combination of light front approach and quark meson coupling model has been used  
 to calculate the in-medium electromagnetic form factors of nucleons \cite{Araujo2018}. 
 In Refs. \cite{harpreet_cpc2017} and 
 \cite{harpreet_epjp2019} chiral quark mean field model has been used to study the magnetic moments of octet baryons  in symmetric and asymmetric nuclear matter, respectively.
 The study of in-medium magnetic moments of baryons has been also extended to symmetric strange  matter for octet \cite{harpreet_epja2018} and decuplet baryons \cite{ harpreet_epjp2020}.
 
As discussed above, various models have been used in the literature to study the magnetic moments of decuplet baryons in the free space, however, the studies in dense matter are  very limited \cite{harpreet_epjp2020,Tsushimaptep2022}. Further, no work is available in our knowledge till date where
the magnetic moments of decuplet baryons have been calculated in isospin asymmetric strange medium.
The aim of the present research paper is to study the influence of medium comprising of nucleons and hyperons with finite isospin asymmetry 
on the masses and magnetic moments of decuplet baryons. The magnetic moments of decuplet baryons have been calculated using the chiral constituent quark model \cite{dahiya2010,aarti} whereas the in-medium effects have been stimulated  through chiral SU(3) quark mean field model \cite{wang_nuc2001,wangstrange2001}. 
In the chiral SU(3) quark mean field model, 
quarks are treated as degrees of freedom and are confined inside baryons through a confining potential. Quarks confined inside the baryons interact through the exchange of
scalar fields $\sigma$, $\zeta$ and $\delta$ and the vector fields $\omega$, $\rho$ and $\phi$.
In Ref. \cite{wangstrange2001}, the properties of isospin symmetric strange matter were investigated using chiral SU(3) quark mean field model considering scalar fields $\sigma$ and $\zeta$ and the vector fields $\omega$ and $\phi$. For the study of isospin asymmetric matter only vector-isovector $\rho$ field was used during  the initial development of this model.  However, as has been emphasized in various studies,
the scalar-isovector meson $\delta$
plays a significant role in the study of isospin asymmetric effects and hence, has been included in the current work \cite{thakur2022,chen2022,Liu2002}.
 Finite baryon density and temperature of the medium lead to the modification in the properties of these constituent quarks and hence the baryons. 
 
 In this context, it would be important to study the properties of decuplet baryons in a medium consisting of nucleons and hyperons with finite isospin asymmetry. The  finite isospin asymmetry of the medium implies the non-zero expectation values of scalar field $\delta$ and vector field $\rho$, whereas finite strangeness fraction in the medium (presence of hyperons along with nucleons)
implies the non-zero value of strange vector field $\phi$. From the chiral SU(3) quark mean field model we shall calculate the medium modified masses of decuplet baryons and their constituent quarks and further use as  them as input in 
the expressions for magnetic moment calculations within the chiral constituent quark model. 

Within the chiral constituent quark model, the magnetic moments of baryons are calculated taking into consideration the contribution of valence quark and sea quark spin polarization
as well as the contribution of orbital angular momentum of sea quarks \cite{dahiya2010,aarti}. As has been emphasized in various studies, the contribution of  sea quark polarization and their orbital angular momentum to the magnetic moments of baryons play an important role in obtaining results in free space comparable to available  experimental information \cite{jan2005,dahiya2010,aarti}.

The present paper is organized as follows: In \cref{sec:chiral_mean_field} we describe the details of the chiral SU(3) quark mean field model. The details of framework used to calculate the magnetic moments of decuplet baryons, i.e., the  chiral constituent quark model is presented in \cref{sec:magnetic}. The results and discussions of the present work are given in \cref{sec:results} and finally the work is summarized  in \cref{sec:summary}.

\section{Chiral SU(3) quark mean field model for asymmetric strange hadronic matter} \label{sec:chiral_mean_field}
In the present work, to study the modification of magnetic moments of baryons in isospin asymmetric strange matter, we use chiral SU(3) quark mean field model which considers quarks and mesons as  degrees of freedom.
As discussed earlier, in the chiral quark mean field model, quarks are confined inside baryons through a confining potential. Chiral SU(3) quark mean field model incorporates the basic ingredients of low energy QCD properties such as chiral symmetry and its spontaneous and explicit breaking \cite{wang_nuc2001}. The constituent quarks of baryons obtain their masses through the exchange of scalar fields $\sigma, \zeta$ and $\delta$.
The effective Lagrangian density of chiral $\text{SU(3)}$ quark mean field model which describes various interaction terms is given by \cite{wang_nuc2001}
\begin{equation}
{\cal L}_{{\rm eff}} \, = \, {\cal L}_{q0} \, + \, {\cal L}_{qm}
\, + \,
{\cal L}_{\Sigma\Sigma} \,+\, {\cal L}_{VV} \,+\, {\cal L}_{\chi SB}\,
+ \, {\cal L}_{\Delta m} \, + \, {\cal L}_{c}. \label{totallag}
\end{equation}
In the above equation,  ${\cal L}_{q0} =\bar q \, i\gamma^\mu \partial_\mu \, q$ represents
kinetic term for free mass-less quarks. The second term,
 ${\cal L}_{qm}$ is the  quark-meson interaction term
 and comprises of interactions of quarks with scalar and vector mesons. Denoting, $\Psi = (u, d, s)$
as the quark field corresponding to  three flavors, the Lagrangian density  ${\cal L}_{qm}$ is written as   \cite{wang_nuc2001,wangstrange2001,harpreet_cpc2017}
\begin{align}
{\cal L}_{qm}=g_s\left(\bar{\Psi}_LM\Psi_R+\bar{\Psi}_RM^{\dagger}\Psi_L\right)
-g_v\left(\bar{\Psi}_L\gamma^\mu l_\mu\Psi_L+\bar{\Psi}_R\gamma^\mu
r_\mu\Psi_R\right)~~~~~~~~~~~~~~~~~~~~~~~  \nonumber \\
=\frac{g_s}{\sqrt{2}}\bar{\Psi}\left(\sum_{a=0}^8 s_a\lambda_a
+ i \gamma^5 \sum_{a=0}^8 p_a\lambda_a
\right)\Psi -\frac{g_v}{2\sqrt{2}}
\bar{\Psi}\left( \gamma^\mu \sum_{a=0}^8
 v_\mu^a\lambda_a
- \gamma^\mu\gamma^5 \sum_{a=0}^8
a_\mu^a\lambda_a\right)\Psi. \label{quarkmesons}
\end{align}
The parameters $g_s$ and $g_v$ appearing in the above equation
are related to the couplings of quarks with scalar and vector mesons fields. The term ${\cal L}_{\Sigma\Sigma}$ and ${\cal L}_{VV}$ appearing
 in \cref{totallag} are the chiral invariant self-interaction terms for scalar and vector mesons, respectively.
 Within the mean field approximation employed in the present work, the self-interaction term for scalar mesons is written as \cite{wang_nuc2001} 
\begin{align}
{\cal L}_{\Sigma\Sigma} =& -\frac{1}{2} \, k_0\chi^2
\left(\sigma^2+\zeta^2+\delta^2\right)+k_1 \left(\sigma^2+\zeta^2+\delta^2\right)^2
+k_2\left(\frac{\sigma^4}{2} +\frac{\delta^4}{2}+3\sigma^2\delta^2+\zeta^4\right)\nonumber \\ 
&+k_3\chi\left(\sigma^2-\delta^2\right)\zeta 
 -k_4\chi^4-\frac14\chi^4 {\rm ln}\frac{\chi^4}{\chi_0^4} +
\frac{\xi}
3\chi^4 {\rm ln}\left(\left(\frac{\left(\sigma^2-\delta^2\right)\zeta}{\sigma_0^2\zeta_0}\right)\left(\frac{\chi^3}{\chi_0^3}\right)\right). \label{scalar0}
\end{align}    
For the vector mesons, we have
\begin{equation}
{\cal L}_{VV}=\frac{1}{2} \, \frac{\chi^2}{\chi_0^2} \left(
m_\omega^2\omega^2+m_\rho^2\rho^2+m_\phi^2\phi^2\right)+g_4\left(\omega^4+6\omega^2\rho^2+\rho^4+2\phi^4\right). 
\label{vector}
\end{equation}
We reiterate that the presence of $\delta$ and $\rho$ mesons
in the above equations  imply that we are investigating the isospin asymmetric matter whereas the presence of $\phi$ in self-interaction terms of vector mesons indicates that the medium we consider is comprising of hyperons along with nucleons.
The last three terms of 
\cref{scalar0} are introduced in the model to incorporate the trace anomaly property of QCD and lead  to the trace of energy momentum tensor proportional to fourth power of dilaton field $\chi$ \cite{papag_1999}.
The order of magnitude about which the value of parameter  $\xi$ is taken into the calculations is generally determined from 
 QCD $\beta$-function at one loop level for three colors and three flavors \cite{papag_1999}. 
 The parameters $k_0, k_1, k_2, k_3$ and $k_4$ appearing in \cref{scalar0} are calculated using $\pi$ meson mass ($m_{\pi}$), $K$ meson mass ($m_K$) and the average mass of $\eta$ and $\eta^{'}$ mesons. 
 
The vacuum expectation values of scalar meson fields $\sigma$ and $\zeta$, i.e., $\sigma_0$ and $\zeta_0$ are expressed in terms of pion decay constant $f_\pi$ and kaon decay constant $f_K$ through relations
\begin{align}
\sigma_0= -f_{\pi} ~~{\rm and}~~~~  \zeta_0= \frac{1}{\sqrt{2}}\left( f_{\pi}-2f_{K}\right).
\end{align} 
In the present work, we consider $f_{\pi}=92.8$ MeV and $f_{K}=115$ MeV whereas the values of $\sigma_0$ and $\zeta_0$ are obtained as $-93.49$ MeV and $-97.98$ MeV  respectively. 
 The vacuum value of dilaton field $\chi_0=254.6$ MeV and the coupling constant $g_4=37.4$ are  fitted to reasonable effective nucleon mass \cite{wang_nuc2001}.
 
 The fifth term of \cref{totallag},
 the Lagrangian density ${\cal L}_{\chi SB}$, is the explicit symmetry breaking term and is written as
\begin{equation}\label{L_SB}
{\cal L}_{\chi SB}=\frac{\chi^2}{\chi_0^2}\left[m_\pi^2f_\pi\sigma +
\left(
\sqrt{2} \, m_K^2f_K-\frac{m_\pi^2}{\sqrt{2}} f_\pi\right)\zeta\right].
\end{equation}
This term is introduced in chiral effective models to embody the non-vanishing pesudoscalar meson masses and it satisfies the partial conserved axial-vector current relations for $\pi$ and $K$ mesons \cite{wang_nuc2001,papag_1999}.

The vacuum masses of constituent quarks (at zero baryon density) 
are expressed in terns of vacuum expectation values $\sigma_0$ and $
$$\zeta_0$ of scalar fields $\sigma$ and $\zeta$, respectively.
For the light $u$ and $d$ quarks, we have
\begin{align}
\label{qvacmass}
m_u=m_d=-g_{\sigma}^q \sigma_0=-\frac{g_s}{\sqrt{2}}\sigma_0.
\end{align}
The values of coupling constant $g_s$ is fitted to obtain
 $m_u=m_d=253$ MeV.
To obtain a reasonable and correct value for the strange quark mass $m_s$, additional mass term $\Delta m$ is introduced through
one more explicit breaking term
in  \cref{totallag} (6th term) and is defined by   \cite{wang_nuc2001,harpreet_cpc2017}
 \begin{equation}
 {\cal L}_{\Delta m} = - (\Delta m) \bar \psi S_1 \psi.
 \end{equation}
Here, the strange quark matrix operator $S_1$ is written as
 \begin{equation}
 S_1 \, = \, \frac{1}{3} \, \left(I - \lambda_8\sqrt{3}\right) =
 {\rm diag}(0,0,1).
 \end{equation}
This leads to following equation for vacuum mass of strange quark
\begin{align}
\label{qvacmass_strange}
m_s=-g_{\zeta}^s \zeta_0 + \Delta m,
\end{align}
where $g_\zeta^s = g_s$ and $\Delta m$ is fitted to obtain $m_s = 450$ MeV.
The last term of \cref{totallag},
corresponding to the confinement of quarks inside the baryons, is written as
\begin{align}
{\cal L}_{c} = -  \bar \psi \chi_c \psi,
\end{align}
where the scalar-vector potential $\chi_c$ is given by \cite{wang_nuc2001}  
\begin{align}
\chi_{c}(r)=\frac14 k_{c} \, r^2(1+\gamma^0) \,.   \label{potential}
\end{align}
The coupling constant $k_c$ is taken to be $98 \, \text{MeV}. \text{fm}^{-2}$.

In order to investigate the properties of asymmetric nuclear matter at finite temperature and density we will use the mean field approximation \cite{wang_nuc2001}. The Dirac equation, under the influence of meson mean field, for the quark field $\Psi_{qi}$, is given as 
\begin{equation}
\left[-i\vec{\alpha}\cdot\vec{\nabla}+\chi_c(r)+\beta m_q^*\right]
\Psi_{qi}=e_q^*\Psi_{qi}, \label{Dirac}
\end{equation}
where the subscripts $q$ and $i$ denote the quark $q$ ($q=u, d, s$)
in a baryon of type $i$ ($i=N, \Lambda, \Sigma, \Xi$)\,
and $\vec{\alpha}$\,, $\beta$\, are the usual Dirac matrices.
The effective quark mass $m_{q}^*$ is defined as
\begin{equation}
m_q^*=-g_\sigma^q\sigma - g_\zeta^q\zeta - g_\delta^q I^{3q} \delta + m_{q0}, \label{qmass}
\end{equation}
where $m_{q0}$ is zero for non-strange `$u$' and `$d$' quarks, whereas for strange `$s$' quark $m_{q0}=\Delta m=77$ MeV. Effective energy of particular quark under the influence of meson field is given as,
$e_q^*=e_q-g_\omega^q\omega-g_\rho^q I^{3q}\rho -g_\phi^q\phi\,$ \cite{wang_nuc2001}.
Various coupling constants used in the present work are related through
\begin{align}
\frac{g_s}{\sqrt{2}}
= &g_{\delta}^u = -g_{\delta}^d = g_\sigma^u = g_\sigma^d  =
\frac{1}{\sqrt{2}}g_\zeta^s, \label{relation}
~~~~~g_{\delta}^s = g_\sigma^s = g_\zeta^u = g_\zeta^d = 0 \, ,\\
&\frac{g_v}{2\sqrt{2}}
= g_{\rho}^u = -g_{\rho}^d = g_\omega^u = g_\omega^d,
~~~~~~~g_\omega^s = g_{\rho}^s  = 0 .
\end{align}
The effective mass of $i^{th}$ baryon is related to the effective energy $E_i^*$ and spurious center of mass momentum $p_{i \, \text{cm}}$ \cite{barik1,barik2} through relation
\begin{align}
M_i^*=\sqrt{E_i^{*2}- <p_{i \, \text{cm}}^{*2}>}. \label{baryonmass}
\end{align} 
The effective energy of baryons $E_i$ is further expressed in terms of effective energy of constituent quarks as 
\begin{equation} 
E_i^*=\sum_qn_{qi}e_q^*+E_{i \, \text{spin}}.
    \label{energy}
\end{equation}
The term $E_{i \, \text{spin}}$ 
contributes as a correction to baryon energy due to spin-spin interaction and is fitted to obtain correct masses of baryons in the free space.
Also, $n_{qi}$ represents the number of quarks of type $q$ in the $i^{th}$ baryon
  
 From the above relations we observe
 that the effective masses of baryons  are related to effective
 quark energy and  masses which in turn are related to in-medium values of vector ($\omega, \rho$ and $\phi$) and scalar  ($\sigma, \zeta$ and $\delta$) fields. 
 To obtain the density and temperature dependent values of scalar and vector fields, we first write the thermodynamic potential in the model for strange isospin asymmetric matter. We have
\begin{equation}
\Omega = -\frac{k_{B}T}
{(2\pi)^3} \sum_{i} \gamma_i
\int_0^\infty d^3k\biggl\{{\rm ln}
\left( 1+e^{- [ E^{\ast}_i(k) - \nu_i^* ]/k_{B}T}\right) \\
+ {\rm ln}\left( 1+e^{- [ E^{\ast}_i(k)+\nu_i^* ]/k_{B}T}
\right) \biggr\} -{\cal L}_{M}-{\cal V}_{\text{vac}}, 
\label{Eq_therm_pot1}  
\end{equation}
where summation is over the nucleons  and hyperons of the medium, i.e., $i = p,n,\Lambda, \Sigma^{\pm,0}, \Xi^{-,0}$.
Also, $
{\cal L}_{M} \, = 
{\cal L}_{\Sigma\Sigma} \,+\, {\cal L}_{VV} \,+\, {\cal L}_{\chi SB}\,
$, with details of individual terms discussed earlier. Also, $\gamma_i=2$ is the degeneracy factor for baryons 
and $E^{\ast }(k)=\sqrt{M_i^{\ast 2}+k^{2}}$. The effective chemical potential $\nu_i^*$ of baryons is related to the free chemical potential $\nu_i$ through relation \cite{wang_nuc2001}
\begin{align}
\nu_i^* = \nu_i - g_{\omega}^i\omega -g_{\rho}^i I^{3i} \rho-g_{\phi}^i\phi.
\end{align}
The thermodynamic potential defined by \cref{Eq_therm_pot1} is minimized with respect to the scalar fields
 $\sigma$, $\zeta$ and $\delta$,
 the dilaton field, $\chi$,  and, the vector fields $\omega$, $\rho$  and $\phi$
 through
 \begin{align}
  \frac{\partial \Omega}{\partial \sigma} = 
  \frac{\partial \Omega}{\partial \zeta} =
  \frac{\partial \Omega}{\partial \delta} =
  \frac{\partial \Omega}{\partial \chi} =
  \frac{\partial \Omega}{\partial \omega} =
  \frac{\partial \Omega}{\partial \rho} =
  \frac{\partial \Omega}{\partial \phi} =
    0.
    \label{eq:therm_min1}
  \end{align}
  The system of non-linear equations
  obtained above (detailed expressions are given in appendix) are solved for different values of baryon density, temperature, isospin asymmetry and strangeness fraction of the medium.
  The finite isospin asymmetry in the medium is introduced through the isospin asymmetry parameter $\eta = -\frac{\Sigma_i I_{3i} \rho_{i}}{\rho_{B}}$, whereas for the finite strangeness fraction, the definition $f_s = \frac{\Sigma_i \vert s_{i} \vert \rho_{i}}{\rho_{B}}$ is used.
  Here,  $I_{3i}$ and $\vert s_{i} \vert$ denote the 3$^{rd}$ component of isospin quantum number and number of strange quarks in $i^{th}$ baryon, respectively. Also, $\rho_B$ is the total baryonic density of the medium.
 
	\section{Magnetic Moment of Decuplet Baryons}
	\label{sec:magnetic}
In the present section we discuss the procedure to calculate the magnetic moments of decuplet baryon
within chiral constituent quark model \cite{dahiya2010,aarti}.
In this model, the internal constituent quark undergoes the emission of Goldstone boson which
future splits into the quark-antiquark pairs
\cite{{chengsu3},{cheng1},{song}}.
The interaction Lagrangian density ${\cal L} = g_8 \bar{q} \Phi_q$  describes the interactions of quarks with Goldstone bosons.  Here, $\Phi$
contain the octet and singlet $\eta^{'}$ and is given by 
\begin{equation}
\Phi=\left(\begin{array}{ccc}\frac{\pi^o}{\sqrt{2}}+\beta \frac{\eta}{\sqrt{6}}+\zeta^{'} \frac{\eta^{\prime}}{4\sqrt{3}} & \pi^{+} & \alpha K^{+} \\ \pi^{-} & -\frac{\pi^o}{\sqrt{2}}+\beta \frac{\eta}{\sqrt{6}}+\zeta^{'} \frac{\eta^{\prime}}{4\sqrt{3}} & \alpha K^0 \\ \alpha K^{-} & \alpha \bar{K}^0 & -\beta \frac{2 \eta}{\sqrt{6}}+\zeta^{'} \frac{\eta^{\prime}}{4\sqrt{3}}\end{array}\right).
\end{equation}
The breaking  of SU(3) symmetry is introduced via constraints  $m_s>m_{u, d}$ and also considering the masses of GBs as nondegenerate $\left(M_{\eta^{\prime}}>M_{K, \eta}>M_\pi\right)$. 
Denoting, parameter $a=\left|g_{8}\right|^2$ as the transition probability of chiral fluctuation of the splitting $u(d) \rightarrow$ $d(u)+\pi^{+(-)}$,  $a \alpha^2, a \beta^2, a \zeta^{'2}$, and $a \gamma^2$ represent the probabilities of transitions of $u(d) \rightarrow s+K^{-(0)}, u(d, s) \rightarrow$ $u(d, s)+\eta$ and $ u(d, s) \rightarrow u(d, s)+\eta^{\prime}$, respectively
 \cite{dahiya2010,aarti}.
 
 As discussed earlier, to calculate the total magnetic moment in the chiral constituent quark model, the contribution of valence  and sea quark spin polarization and orbital angular momentum of sea quark is considered.
We can write the equation
\begin{align}
	 \mu_{B}^*= \mu_{\text{B,val}}^*+\mu_{\text{B,sea}}^*+\mu_{\text{B,orbital}}^*, \label{magtotal}
	 \end{align}
	 where $\mu_{\text{B, val}}^*$, $\mu_{\text{B, sea}}^*$ and $\mu_{\text{B, orbital}}^*$ represent the contribution from valence quarks, sea quarks and orbital angular momentum of sea quarks, respectively. 
	 The asteric sign here implies that we will calculate these in the strongly interacting medium where these values will be different compared to free space.
Individually, these contribution are
calculated using expressions
\begin{align}
\mu_{\mathrm{B,val}}^* & =\sum_{q=u, d, s} \Delta q_{\mathrm{val}} \mu_q^*, 
\label{mu_val1}
\\
\mu_{\mathrm{B,sea}}^* & =\sum_{q=u, d, s} \Delta q_{\mathrm{sea}} \mu_q^*,
\label{mu_sea1} \\
\mu_{\text {B,orbit }}^* & =\sum_{q=u, d, s} \Delta q_{\mathrm{val}} \mu^*\left(q_{+} \rightarrow q_{-}^{\prime}\right).
\label{mu_seaorbit1}
\end{align}
In above equations, 
$\Delta q_{\mathrm{val}}$ and 
$\Delta q_{\mathrm{sea}}$ represent the
spin polarization due to valance and sea quarks.
In \cref{mu_val1,mu_sea1} $\mu_q^*$ is the magnetic moments of quarks in the strange hadronic
 medium.
 Although the value of $\mu_q^*$  can be calculated using formula 
 	$\mu_{\rm q}^*=\frac{e_{\rm q}}{2m_{\rm q}^*}$, where $m_{\rm q}^*$ and $e_{\rm q}$ are mass and electric charge of quark, respectively, however, this formula lacks consistency for calculation of magnetic moments of relativistically confined quarks \cite{aarti}.
%
 The impact of quark confinement and relativistic corrections can be more correctly incorporated using the following expressions for the effective magnetic moments of constituent quarks of a given baryon	
 	\begin{equation}
 	 \mu_d^* =-\left(1-\frac{\Delta M}{M_B^*}\right),~~  \mu_s^*=-\frac{m_u^*}{m_s^*}\left(1-\frac{\Delta M}{M_B^*}\right),~~  \mu_u^*=-2\mu_d^* .\label{magandmass}         
 	\end{equation}     
 	These 
 	are referred to as the mass adjusted magnetic moments of constituent quarks \cite{aarti}.
 	Here, $M_B^*$ is the effective mass of baryon whose magnetic moment we wish to calculate
 	and is obtained using \cref{baryonmass}. Also, 
 	$\Delta M=M_{B}^{*} - M_{\rm vac}$, where $M_{\rm vac}$ is the mass of baryon in the free space.
 	In \cref{mu_seaorbit1}, $\mu^*\left(q_{+} \rightarrow q_{-}^{\prime}\right)$ represents the orbital moment for chiral fluctuation
 	which are given by  \cite{dahiya2010}
 	\begin{align}
 {\left[\mu^*\left(u_{+} \rightarrow\right)\right]=} & a\left[\frac{3 m_u^{*2}}{2 M_\pi\left(m_u^{*}+M_\pi\right)}-\frac{\alpha^2\left(M_K^2-3 m_u^{*2}\right)}{2 M_K\left(m_u^{*}+M_K\right)}\right. \nonumber\\ & +\frac{\beta^2 M_\eta}{6\left(m_u^{*}+M_\eta\right)} 
  \left.+\frac{\zeta^{'2} M_{\eta^{\prime}}}{48\left(m_u^{*}+M_{\eta^{\prime}}\right)}\right] \mu_N, 
 	\end{align} 
 	
\begin{align}
{\left[\mu^*\left(d_{+} \rightarrow\right)\right]=} & a \frac{m_u^*}{m_d^*}\left[\frac{3\left(M_\pi^2-2 m_d^{*2}\right)}{4 M_\pi\left(m_d^{*2}+M_\pi\right)}-\frac{\alpha^2 M_K}{2\left(m_d^*+M_K\right)}\right.\nonumber \\ & -\frac{\beta^2 M_\eta}{12\left(m_d^*+M_\eta\right)}  \left.-\frac{\zeta^{'2} M_{\eta^{\prime}}}{96\left(m_d^*+M_{\eta^{\prime}}\right)}\right] \mu_N,
\end{align}

\begin{align}
 {\left[\mu^*\left(s_{+} \rightarrow\right)\right]=} & a \frac{m_u^*}{m_s^*}\left[\frac{\alpha^2\left(M_K^2-3 m_s^{*2}\right)}{2 M_K\left(m_s^*+M_K\right)}  -\frac{\beta^2 M_\eta}{3\left(m_s^*+M_\eta\right)} \right. \nonumber\\& \left.-\frac{\zeta^{'2} M_{\eta^{\prime}}}{96\left(m_s^*+M_{\eta^{\prime}}\right)}  \right] \mu_N,
\end{align} 
where $\mu_N$ is nuclear magneton.
As can be see from above equations,
the orbital moment for chiral fluctuatons, $\mu^*\left(q_{+} \rightarrow q_{-}^{\prime}\right)$ 
is modified in the strange isospin asymmetric matter through the medium  modified masses of constituent quarks.
%
 	

	\section{Numerical results} \label{sec:results}
In this section we present in detail
the results of medium modification of masses and magnetic moments of decuplet baryons in strange hadronic medium with finite isospin asymmetry
and temperature.
As has been discussed earlier, the in-medium masses of baryons in the strange matter are calculated using medium modified  constituent quark masses and effective energies which in turn 
are calculated using scalar fields
$\sigma, \zeta$ and $\delta$ and  vector fields $\omega, \rho$ and $\phi$. Values of various parameters are given in \cref{table1_para}.
The coupling  of hyperons to the scalar fields are calculated using following relations
\begin{align}
g_\sigma^\Lambda &= 2g_\sigma^u + g_\sigma^s; \qquad
g_\zeta^\Lambda = 2g_\zeta^u+g_\zeta^s; \qquad 
g_\delta^\Lambda = 0 \nonumber\\
g_\sigma^\Sigma &= 2g_\sigma^u + g_\sigma^s; \qquad
g_\zeta^\Sigma = 2g_\zeta^u+g_\zeta^s; \qquad 
g_\delta^\Sigma = g_\delta^p \nonumber\\
g_\sigma^\Xi &= g_\sigma^u + 2g_\sigma^s; \qquad
g_\zeta^\Xi = g_\zeta^u+2g_\zeta^s; \qquad 
g_\delta^\Xi = g_\delta^p  
\end{align}
Coupling of  hyperons with vector mesons are fitted using relations 
\begin{align}
g_\omega^\Lambda = x_\Lambda \frac{2}{3}g_\omega^N;\qquad
g_\rho^\Lambda = 0; \qquad
g_\phi^\Lambda=-\frac{\sqrt{2}}{3}g_\omega^N \nonumber\\
g_\omega^\Sigma = x_\Sigma \frac{2}{3}g_\omega^N;\qquad
g_\rho^\Sigma = \frac{2}{3}g_\omega^N;
 \qquad
g_\phi^\Sigma=-\frac{\sqrt{2}}{3}g_\omega^N \nonumber\\
g_\omega^\Xi = x_\Xi \frac{1}{3}g_\omega^N;\qquad
g_\rho^\Xi = \frac{2}{3}g_\omega^N;
 \qquad
g_\phi^\Xi=-\frac{2\sqrt{2}}{3}g_\omega^N \nonumber\\
\end{align}
The parameters $ x_\Lambda,  x_\Sigma$ and  $x_\Xi$ appearing in the above equations are fitted to obtain the reasonable hyperon potential values at nuclear saturation density of nuclear matter.
The values of hyperons potentials are fitted to $U_\Lambda = -30$ MeV,
$U_\Sigma = 30$ MeV and $U_\Xi = -18$ MeV   \cite{millener2001,yamamoto1988,
  mares1995,bart1999,fukuda1998}. Coupling of $\phi$ meson with nucleons is taken as zero, i.e., $g_\phi^N = 0$.
We divide further discussion of this section into two subsections. 
In  \cref{sec_quark_de_mass},
we discuss the medium modified masses of constituent quarks and baryons.
The medium modified masses of quarks and decuplet baryons will be used as input to calculate the in-medium magnetic moments of decuplet baryons which is discussed in \cref{sec_deculet_moments}.
\subsection{In-medium quark and decuplet baryon masses}
\label{sec_quark_de_mass}
\begin{figure}[ht!] 
 \includegraphics[width= 13 cm, height=16 cm]{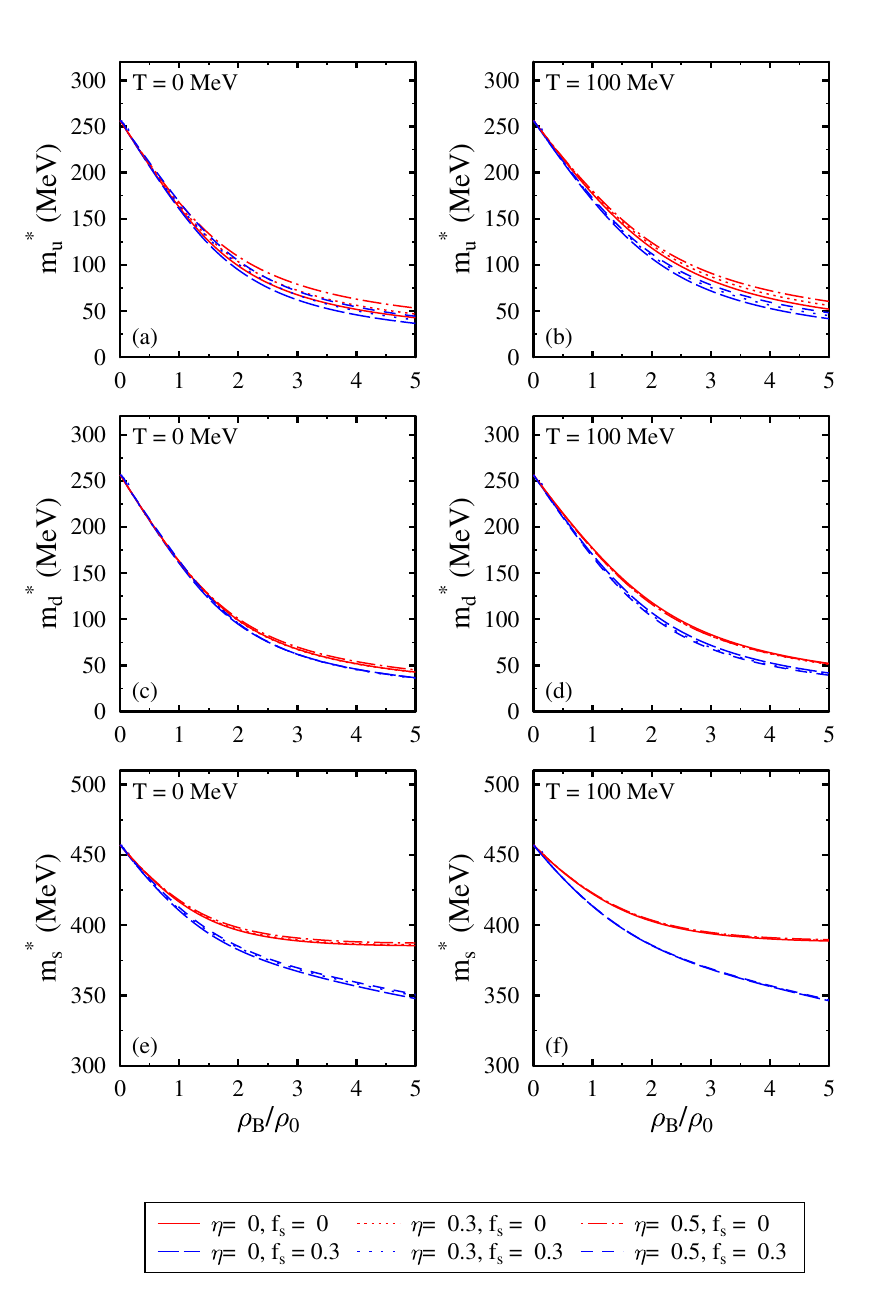}\hfill
	\caption{Effective masses of constituent quarks $u, d$ and $s$ are shown as a function of baryon density $\rho_B$ (in units of nuclear saturation density $\rho_0$) at temperatures $T = 0$ and 100 MeV. Results are shown for isospin asymmetry $\eta = 0, 0.3, 0.5$ and strangeness fractions $f_s = 0$ and $0.3$.}
	\label{fig_massquarks}
\end{figure}

In \cref{fig_massquarks}, we have shown the variation of in-medium constituent
quark masses $m_u^*, m_d^*$ and $m_s^*$  as a function of baryon density $\rho_B$ (in units of nuclear saturation density $\rho_0$), for temperatures $T = 0$ (left panel) and $100$ MeV (right panel).
In each subplot, results are shown for isospin asymmetry $\eta = 0, 0.3, 0.5$ and for each of these values of $\eta$,  the strangeness fraction
is fixed at $f_s = 0$ and $0.3$.

For given temperature $T$, isospin asymmetry $\eta$ and strangeness fraction $f_s$ of the medium, the effective masses of quarks are observed to decrease with an increase in the baryon density $\rho_B$ of the medium, with sharper decrease at lower densities. The density effects are observed to be more appreciable for 
light $u$ and $d$ quarks as compared to strange $s$ quark. For example, in symmetric nuclear matter ($\eta = 0$ and $f_s = 0$), at baryon saturation density $\rho_B =\rho_0$, effective masses of $u$ and $s$ quarks decrease by $36.6$ \% and $8.85$ \%, from their free space values ($\rho_B = T = 0$).
However, when hyperons are included in the medium along with nucleons, the effective mass of strange quark also decreases significantly. Finite strangeness fraction also leads to further decrease in the masses of light $u$ and $d$ quarks. In the symmetric strange medium, with strangeness fraction $f_s =0.3$, the effective masses of $u$ and $s$ quarks decrease by $37.38$ \% and $10.21$ \% from  vacuum values, at baryon density $\rho_B = \rho_0$ and $T = 0$.  The impact of strangeness fraction on the mass modification of strange $s$ quarks is higher in comparison to light $u$ and $d$ quarks at high $\rho_B$.
For example, at baryon density $\rho_B = 3\rho_0$ and temperature $T =0$, in the symmetric medium, as the strangeness fraction $f_s$ is
increased from zero to $0.3$, 
the mass of $u$ quark decreases from value $67.63$ to $62.39$ MeV ($5.24$ MeV drop) whereas the effective mass of $s$ quark decreases from $389$ to $367$ MeV ($22$ MeV drop).

The finite isospin asymmetry of the medium is observed to cause the mass splitting among the light quark isospin doublet ($u$ and $d$ quarks).
This can be realized from \cref{qmass}
where the presence of $\delta$ meson term causes the mass splitting in the masses of light quarks. As discussed earlier, the scalar-isovector meson $\delta$ has a non-zero value in the isospin asymmetric  matter
and since $I^{3u} = 1/2 = -I^{3d}$, this term causes the splitting between the masses of $u$ and $d$ quarks. For a given density, temperature and strangeness fraction of the medium, the increase of isospin asymmetry parameter from $\eta = 0$ to $0.3$ causes a decrease in the effective mass of $d$ quark and an increase in the mass of $u$ quark. Because of zero coupling of $s$ quark with $\delta$ field, the impact of increasing the value of isospin asymmetry is more on light $u$ and $d$ quarks compared to that on the strange $s$ quark. At zero temperature and for baryon density $\rho_B = 3\rho_0$ and strangeness fraction $f_s = 0$,
as isospin asymmetry is increased from $\eta = 0$ to $0.3$,  effective mass of $u$ quark increases from $67.63$ to $72.43$ MeV, whereas the mass of $d$ quark decreases from $67.63$ to $67.26$ MeV.
At strangeness fraction $f_s = 0.3$,
the mass of $u$ ($d$) increases (decreases) by $5.08$ (0.42)  MeV  
as $\eta$ is changed from $0$ to $0.3$.  Finite isospin asymmetry causes a decrease in the impact of strangeness fraction (which was causing decrease in effective mass). In the hadronic medium, the increase of temperature
causes an increase in the effective mass of quarks.

\begin{figure}[ht!] 
 \includegraphics[width= 13 cm, height=18 cm]{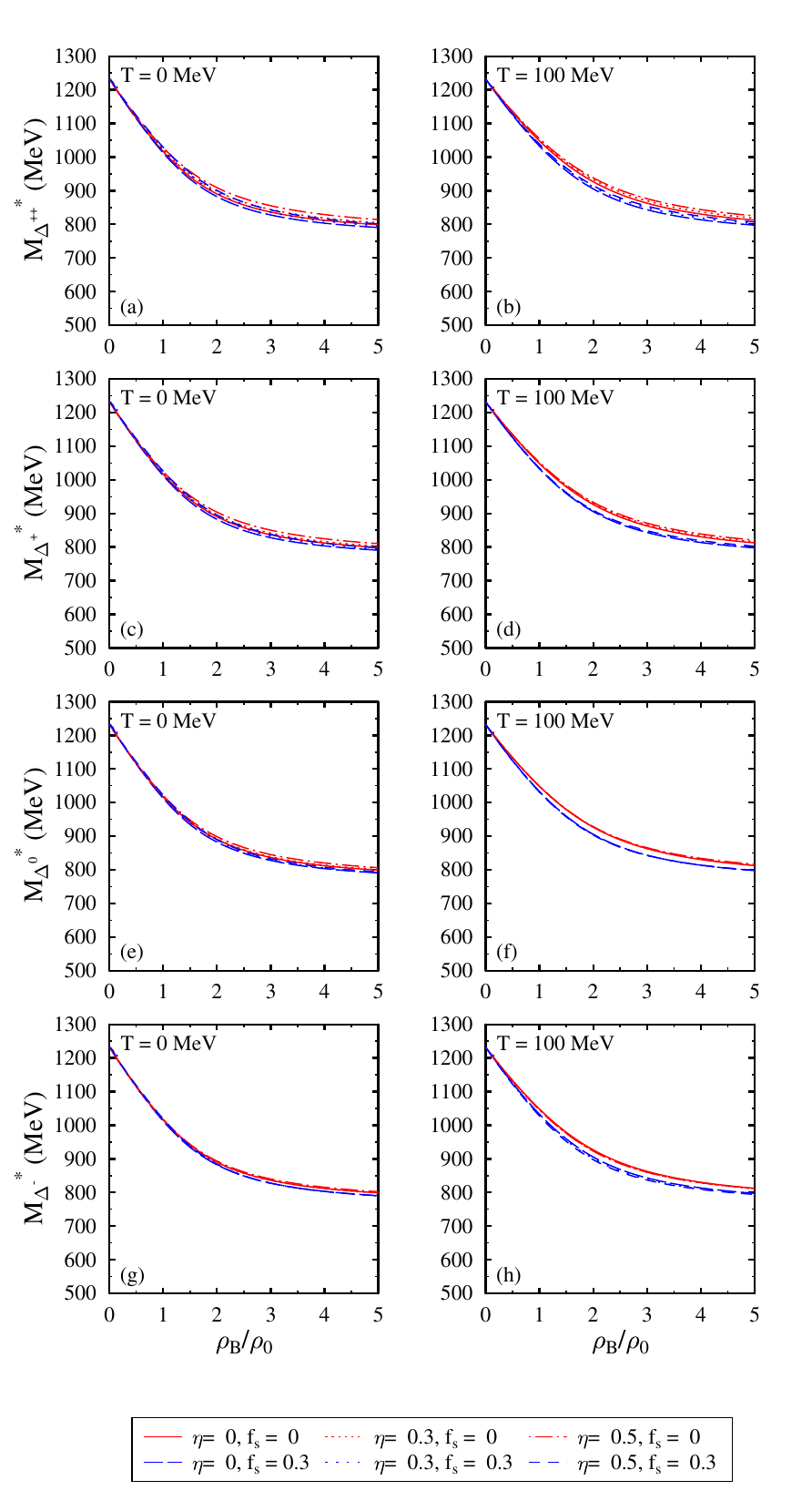}\hfill
	\caption{Effective masses of $\Delta$ resonance baryons $\Delta^{++}, \Delta^+, \Delta^0$ and $\Delta^-$ are shown as a function of baryon density $\rho_B$ (in units of nuclear saturation density $\rho_0$) at temperatures $T = 0$ and 100 MeV. Results are shown for isospin asymmetry $\eta = 0, 0.3, 0.5$ and strangeness fractions $f_s = 0$ and $0.3$.}
	\label{fig_massDelta}
\end{figure}

\begin{figure}[ht!] 
 \includegraphics[width= 15 cm, height=20 cm]{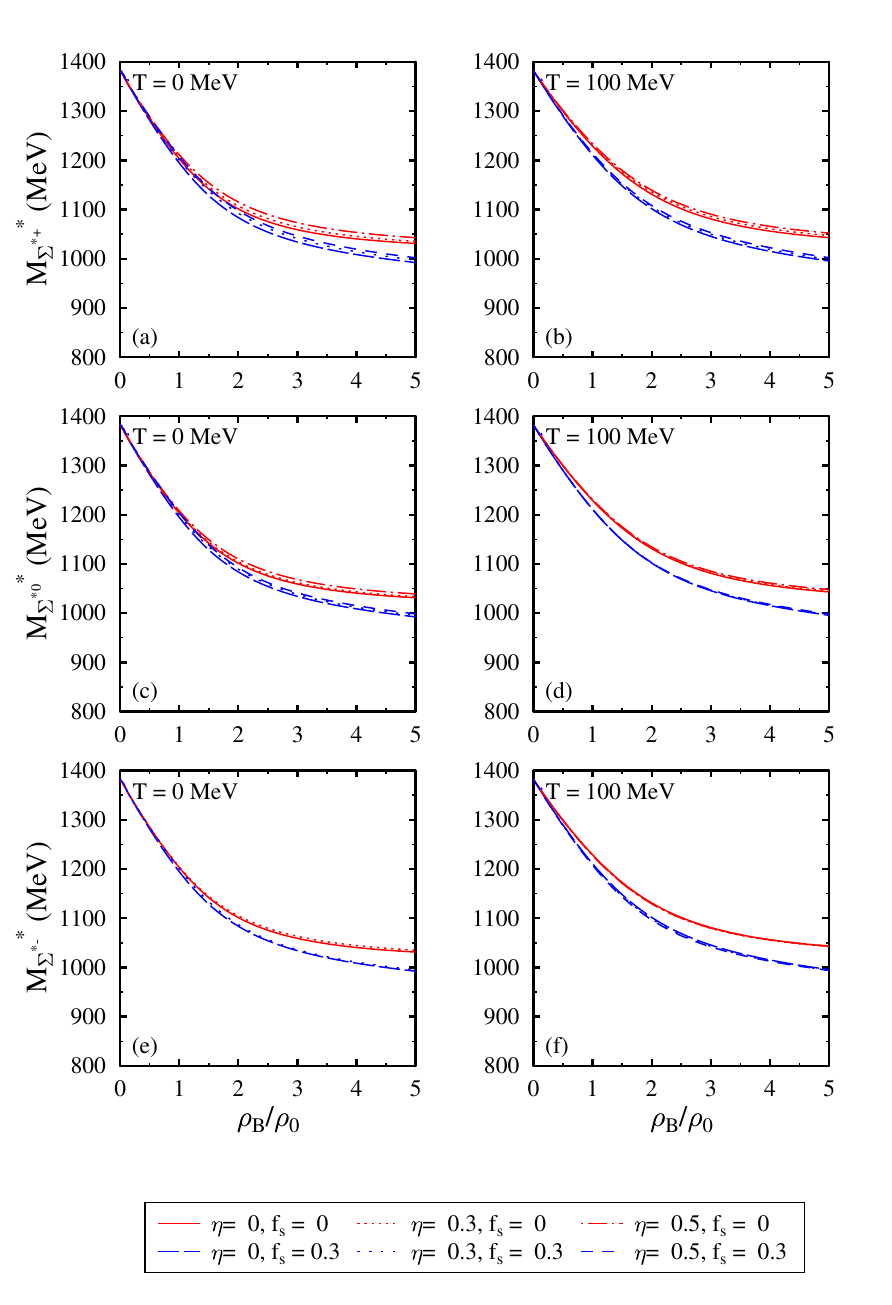}\hfill
	\caption{Effective masses of $\Sigma^*$ resonance baryons $\Sigma^{*+}, \Sigma^{*0}$ and $\Sigma^{*-}$ are shown as a function of baryon density $\rho_B$ (in units of nuclear saturation density $\rho_0$) at temperatures $T = 0$ and 100 MeV. Results are shown for isospin asymmetry $\eta = 0, 0.3, 0.5$ and strangeness fractions $f_s = 0$ and $0.3$.}
	\label{fig_massSigma}
\end{figure}

\begin{figure}[ht!] 
 \includegraphics[width= 13 cm, height=18 cm]{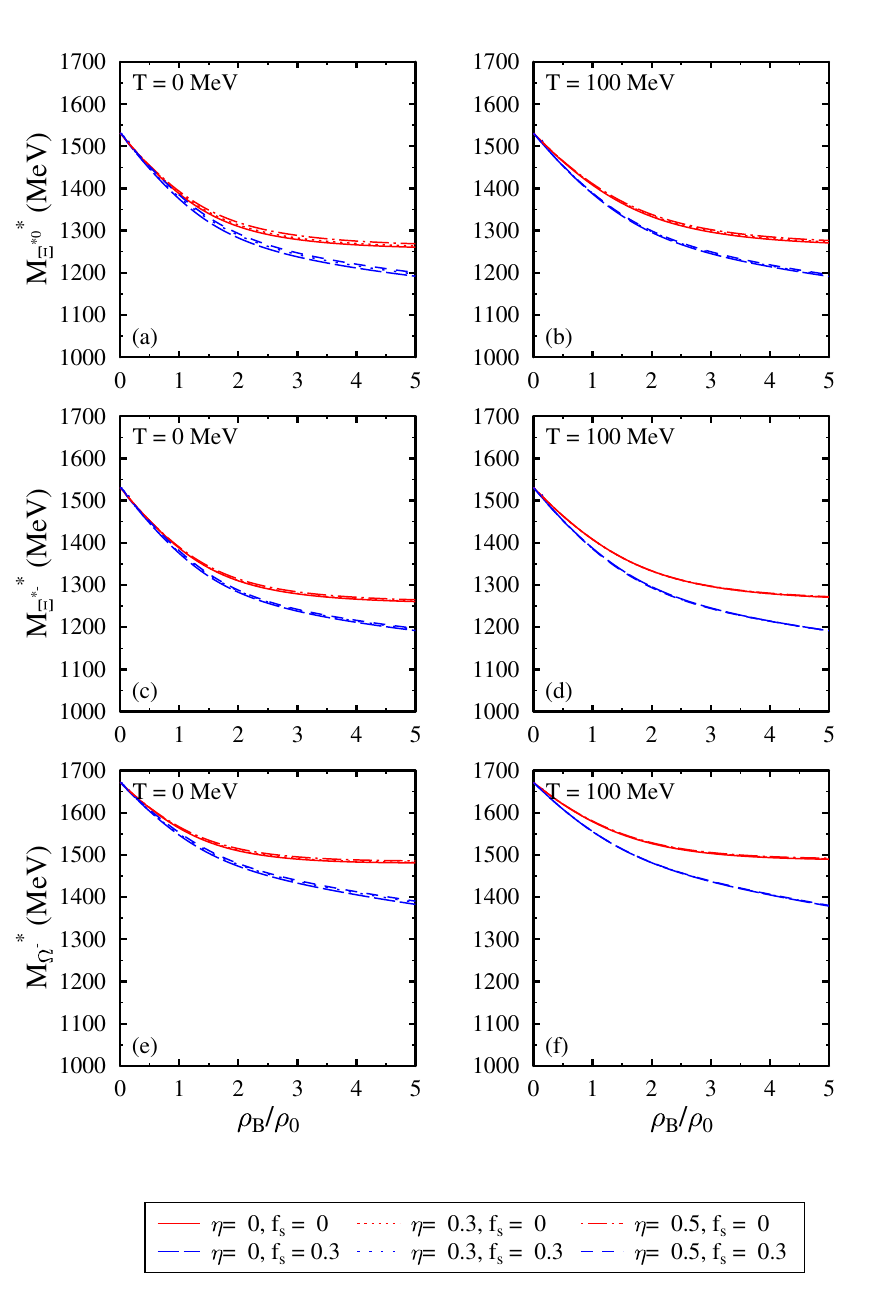}\hfill
	\caption{Effective masses  resonance baryons $\Xi^{-}, \Xi^0$ and $\Omega^-$ are shown as a function of baryon density $\rho_B$ (in units of nuclear saturation density $\rho_0$) at temperatures $T = 0$ and 100 MeV. Results are shown for isospin asymmetry $\eta = 0, 0.3, 0.5$ and strangeness fractions $f_s = 0$ and $0.3$.}
	\label{fig_massXiomega}
\end{figure}

Using the effective constituent quark masses and effective quark energies,  the effective masses of decuplet baryons are calculated in the isospin asymmetric strange matter using
\cref{baryonmass}. The values of $E_{i \text{spin}}$ (\cref{energy}) is fitted to obtain the vacuum masses of decuplet baryons.  In 
\cref{fig_massDelta,fig_massSigma,fig_massXiomega}, we have shown the variation of in-medium masses of decuplet baryons $\Delta$, $\Sigma^{\pm,0}$,  $\Xi^{-,0}$ and $\Omega^{-}$ baryons.
Behaviour of in-medium constituent quarks masses is reflected in the trends of baryon masses.
Among the different decuplet baryons, the 
masses of $\Delta$ baryons decrease more appreciably as compared to other baryons. This is because of light quark content ($u$ and $d$ quarks) of $\Delta$ baryons, whereas the others have one or more strange $s$ quarks present. Among the different $\Delta$ baryons, for given $\rho_B$, $T$ and $f_s$,   the effective mass of $\Delta^{++}$ baryon, which is composed of three $u$ quarks increases with isospin asymmetry of the medium. As the number of $d$ quark increases, i.e., 
in moving from $\Delta^{++}$
to $\Delta^{-}$, the increase in the mass of these baryons becomes slower and for  $\Delta^{-}$ ($ddd$ quarks) it shows an opposite behavior, i.e., decreases with increase in value of  $\eta$. In symmetric medium, at baryon density $3\rho_0$ and strangeness fraction $f_s = 0(0.3)$, the effective mass of $\Delta$ baryons is observed to be
836.54 (827.82) MeV. As isospin asymmetry is increased to finite non-zero value, splitting in the masses of different members of $\Delta$ baryon multiplet takes place.
In asymmetric matter with $\eta = 0.3$ and temperature $T = 0$, the effective masses of $\Delta^{++}, \Delta^{+}, \Delta^{0}$ and $\Delta^{-}$ for $f_s = 0 (0.3)$ are observed to be 844 (836.28), 841.55 (833.37), 838.75 (830.46) and 835.96 (827.55) MeV, respectively.
Increase of temperature of medium from $T = 0$ to 100 MeV causes suppressed decrease in the mass of decuplet baryons.
As can be seen from
\cref{fig_massSigma,fig_massXiomega}, increase of strangeness fractions causes significant decrease in the mass of baryons having more number of strange quarks. For example, in symmetric matter with $\eta = 0$, at $\rho_B = 3\rho_0$, increase of $f_s$ from zero to $0.3$, causes a decrease of $9.34, 25, 41$  and $57$ MeV in the effective masses  of $\Delta$, $\Sigma$, $\Xi$ and $\Omega$ baryons, respectively.
As observed for $\Delta$ baryons, in case of $\Sigma$ baryons also, as one moves from $\Sigma^{+}$ to $\Sigma^{-}$, the increase of isospin asymmetry causes a decrease  in the in-medium mass  of baryons since quark content changes from $u$ to $d$ quarks.

\subsection{In-medium magnetic moments of decuplet baryons}
\label{sec_deculet_moments}
\begin{figure}[ht!] 
 \includegraphics[width= 13 cm, height=18 cm]{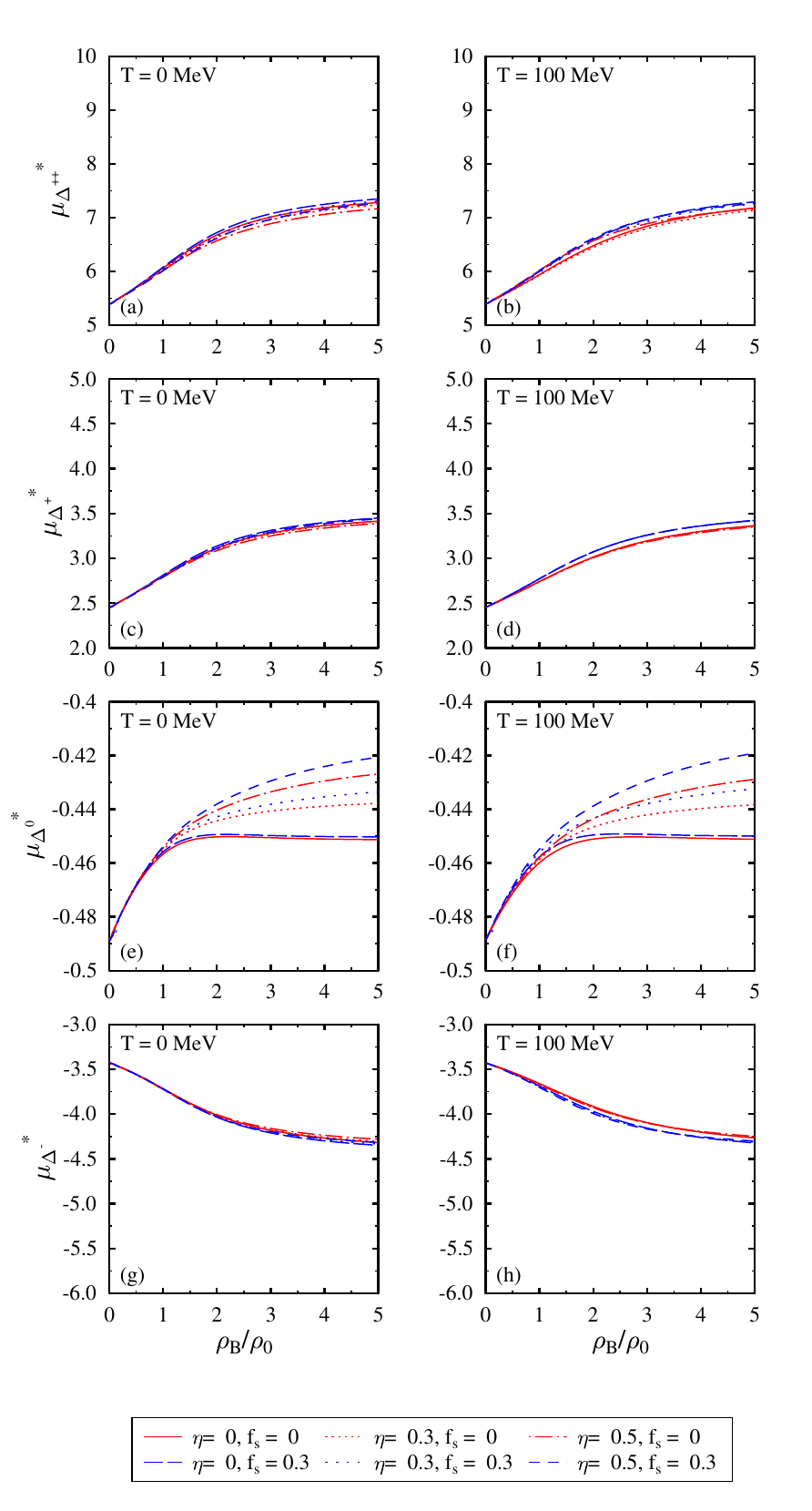}\hfill
	\caption{In-medium magnetic moment $\mu_\Delta^*$ of $\Delta$ resonance baryons ($\mu_{\Delta}^{*++}, \mu_{\Delta}^{*+}, \mu_{\Delta}^{*0}$ and $\mu_{\Delta}^{*-}$)
	is shown as a function of baryon density $\rho_B$ (in units of nuclear saturation density $\rho_0$) at temperatures $T = 0$ and 100 MeV. Results are shown for isospin asymmetry $\eta = 0, 0.3, 0.5$ and strangeness fractions $f_s = 0$ and $0.3$.}
	\label{fig_moment_Delta}
\end{figure}

\begin{figure}[ht!] 
 \includegraphics[width= 17 cm, height=18 cm]{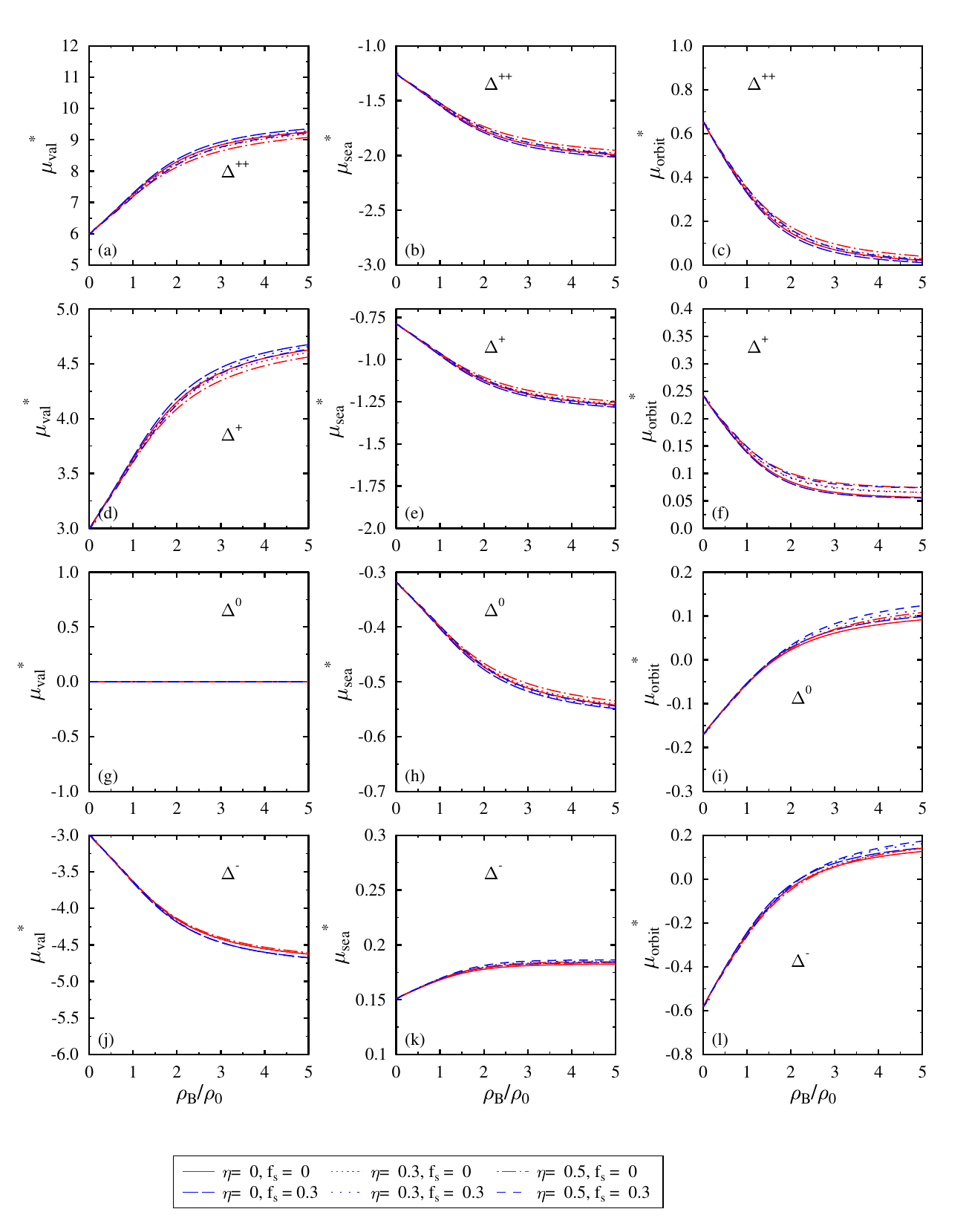}\hfill
	\caption{Contribution to total magnetic moment of
	$\Delta$ resonance baryons from   valance
	quarks, sea quarks and orbital moment of sea quark magnetic moments ($\mu_{val}^{*}$,$\mu_{sea}^{*}$ and $\mu_{orbit}^{*}$, respectively)   
	is shown as a function of baryon density $\rho_B$ (in units of nuclear saturation density $\rho_0$) at temperature $T = 0$ . Results are shown for isospin asymmetry $\eta = 0, 0.3, 0.5$ and strangeness fractions $f_s = 0$ and $0.3$.}
	\label{fig_moment_Delta_indi}
\end{figure}

In the present section, we discuss the results on the medium modification of  magnetic moments of decuplet baryons. The values of magnetic moments of decuplet 
baryons are tabulated in \cref{table:mub_eta0fs0,table:mub_eta5fs0,table:mub_eta0fs3,table:mub_eta5fs3} 
In \cref{fig_moment_Delta}, we have shown the variation of effective magnetic moment $\mu_{\Delta}^{*}$ of $\Delta$ baryons ($\Delta^{++}, \Delta^{+}, \Delta^{0}$ and $\Delta^{-}$) as a function of baryon density $\rho_B$ (in units of nuclear saturation density $\rho_0$).
The magnetic moments of $\Delta^{++}$ and $\Delta^{+}$ baryons increase with baryon density of the medium, whereas for $\Delta^{-}$, it decreases.
In case of $\Delta^{0}$, at $\eta = 0$,  $\mu_{\Delta^{0}}^{*}$ initially increases, reaches to a maximum and then decreases slightly with the further increase in $\rho_{B}$.  However, at finite $\eta$, the magnetic moment of 
$\Delta^{0}$ increases with increase in baryon density $\rho_B$.
Since  the total magnetic moment of baryons in the present work includes the contribution of valence quarks, sea quarks and orbital moment of sea quarks, the trend of total magnetic moment of a given baryon in the medium is therefore determined by the behaviour of these individual contributions. To understand better, in \cref{fig_moment_Delta_indi} we have  shown the variation of these individual magnetic moment contributions as a function of density.
As one can see, in case of $\Delta^{++}$ and $\Delta^{+}$ baryons, the valance quark magnetic moment $\mu_{val}^{*}$ increase with $\rho_B$ whereas $\mu_{sea}^{*}$ 
and $\mu_{orbit}^{*}$ decrease. As the increase of $\mu_{val}^{*}$ dominates over the decrease caused by sea quark and orbital moment of sea quarks, and therefore, total magnetic moment $\mu_{\Delta^{++}}^{*}$ and $\mu_{\Delta^{+}}^{*}$ increase as a function of $\rho_B$. On the otherside,  for $\Delta^{-}$ baryon,
$\mu_{val}^{*}$ decreases as a function of $\rho_B$ and dominates over the increase caused due to
$\mu_{sea}^{*}$ and $\mu_{orbit}^{*}$
(subplots (j), (k) and (l) of \cref{fig_moment_Delta_indi}).
The contribution of valence quarks to the magnetic moment of 
$\Delta^{0}$ remain zero as can be seen form subplot (g) of \cref{fig_moment_Delta_indi}. 
The  magnetic moment due to sea quarks  $\mu_{sea}^{*}$
decreases with $\rho_B$ whereas $\mu_{orbit}^{*}$ increases (sublots (h) and (k)). At $\eta = 0$, initially as density is increased from zero to finite value, the 
$\mu_{orbit}^{*}$ dominates over $\mu_{sea}^{*}$  and therefore, 
$\mu_{\Delta^{0}}$ increases upto certain value. For further increase in $\rho_B$, the $\mu_{sea}^{*}$ starts dominating and causes overall slight decrease in the magnetic moment of $\Delta^0$ baryon as a function of baryon density $\rho_B$.

For fixed value of baryon density $\rho_{B}$, isospin asymmetry $\eta$ and temperature $T$,  increase of strangeness fraction in the medium causes an increase in the magnetic moment of $\Delta^{++}$ and $\Delta^{+}$ baryons and a decrease for $\Delta^{0}$ and $\Delta^{-}$.
An increase in the isospin asymmetry of the medium, i.e, changing $\eta$ from
$0$ to $0.5$ causes a decrease in the
magnetic moment of 
$\Delta^{++}$, $\Delta^{+}$ and $\Delta^{-}$ baryons whereas 
an increase in the value of  magnetic moment for $\Delta^{0}$.
As can be seen from subplots (h) and (i) of \cref{fig_moment_Delta_indi},
the increase of isospin asymmetry in the medium
causes an increase in the value of $\mu_{sea}^{*}$ and $\mu_{orbit}^{*}$. 
 Thus, as shown in subplots (e) and  (f) of \cref{fig_moment_Delta}, at finite isospin asymmetry of the medium the magnetic moment values  of $\Delta^{0}$ increases as a  function of density of the medium.

\begin{figure}[ht!] 
 \includegraphics[width= 13 cm, height=18 cm]{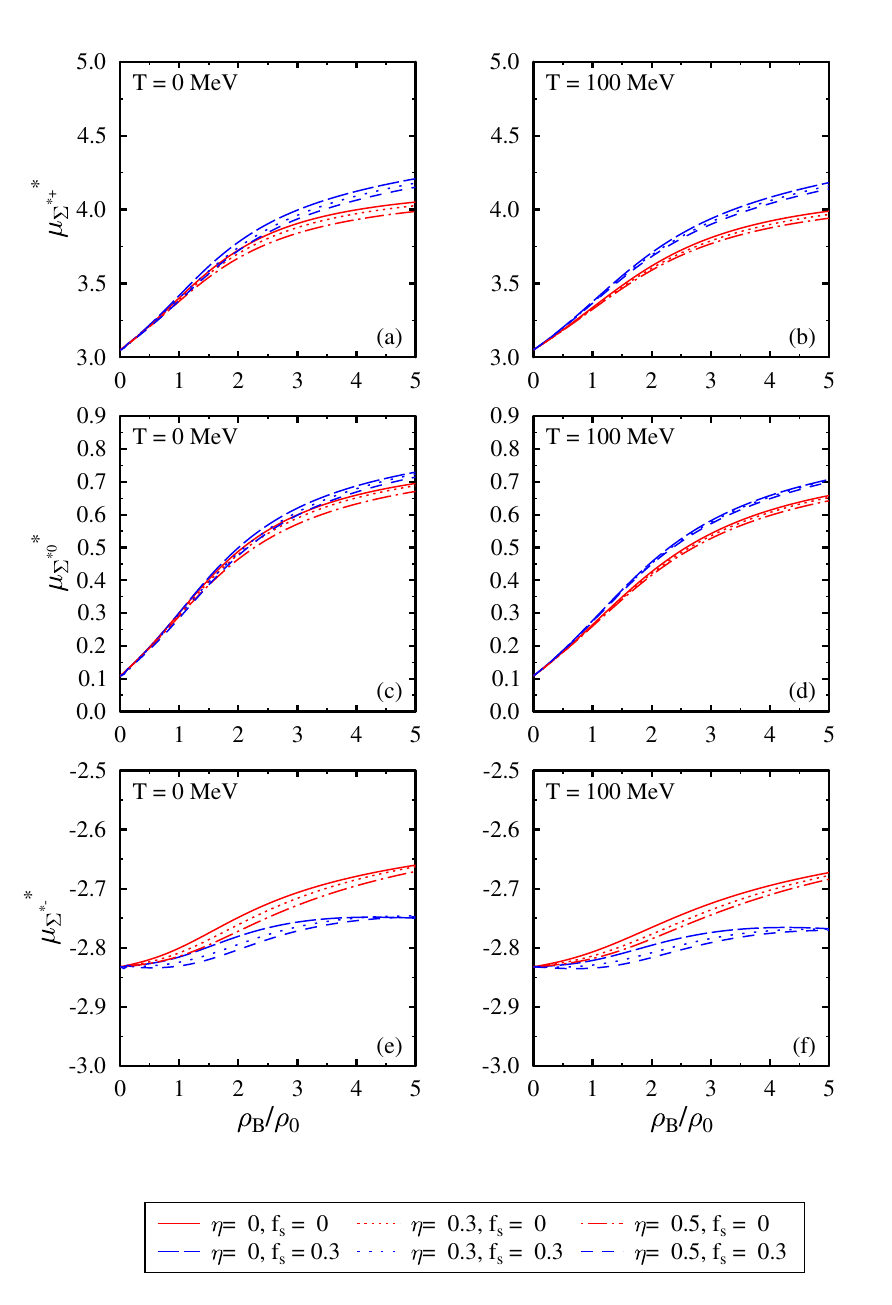}\hfill
	\caption{In-medium magnetic moment $\mu_\Sigma^*$ of $\Delta$ resonance baryons ($\mu_{\Sigma}^{*+},  \mu_{\Sigma}^{*0}$ and $\mu_{\Sigma}^{*-}$)
	is shown as a function of baryon density $\rho_B$ (in units of nuclear saturation density $\rho_0$) at temperatures $T = 0$ and 100 MeV. Results are shown for isospin asymmetry $\eta = 0, 0.3, 0.5$ and strangeness fractions $f_s = 0$ and $0.3$.}
	\label{fig_moment_Sigma}
\end{figure}

\begin{figure}[ht!] 
 \includegraphics[width= 17 cm, height=18 cm]{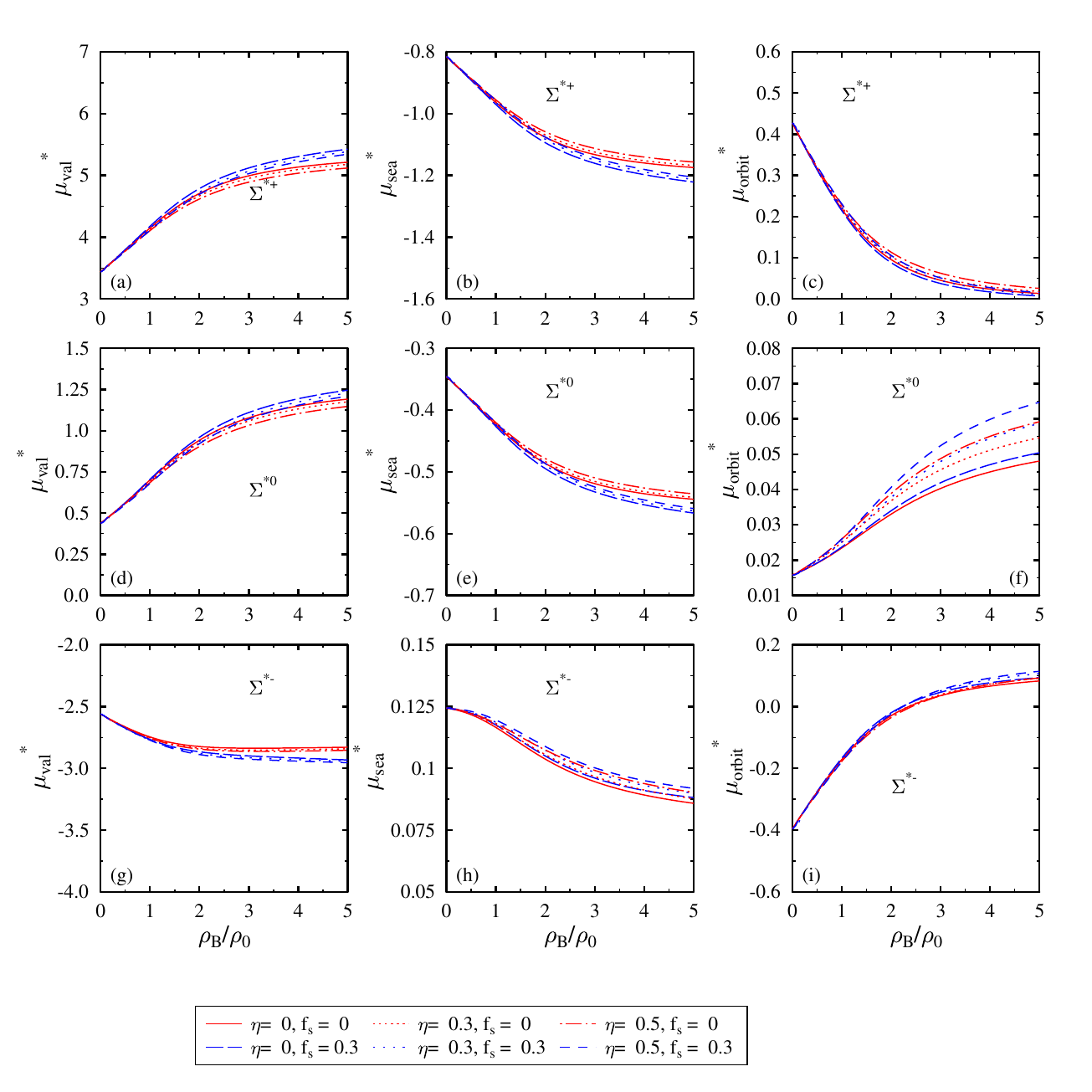}\hfill
	\caption{Contribution to total magnetic moment of
	$\Sigma^{*}$ resonance baryons from   valance
	quarks, sea quarks and orbital moment of sea quark magnetic moments ($\mu_{val}^{*}$,$\mu_{sea}^{*}$ and $\mu_{orbit}^{*}$, respectively)   
	is shown as a function of baryon density $\rho_B$ (in units of nuclear saturation density $\rho_0$) at temperature $T = 0$ . Results are shown for isospin asymmetry $\eta = 0, 0.3, 0.5$ and strangeness fractions $f_s = 0$ and $0.3$.}
	\label{fig_moment_Sigma_indi}
\end{figure}

\begin{figure}[ht!] 
 \includegraphics[width= 13 cm, height=18 cm]{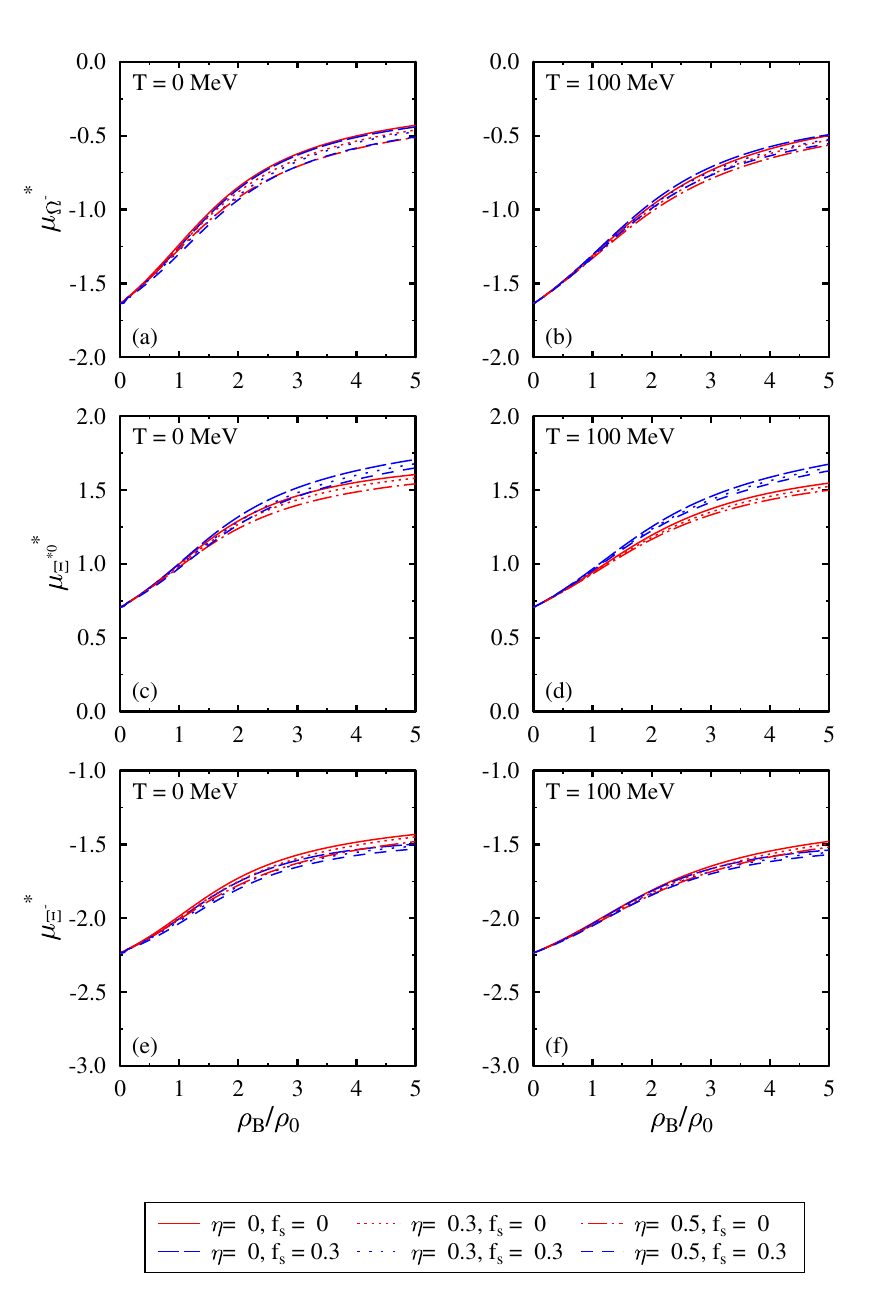}\hfill
	\caption{In-medium magnetic moment $\mu_{\Xi^*}^{*}$ ($\mu_{\Xi^{*0}}^{*}$ and $\mu_{\Xi^{*-}}^{*}$) and
	$\mu_{\Omega^-}^{*}$   resonance baryons  
	is shown as a function of baryon density $\rho_B$ (in units of nuclear saturation density $\rho_0$) at temperatures $T = 0$ and 100 MeV. Results are shown for isospin asymmetry $\eta = 0, 0.3, 0.5$ and strangeness fractions $f_s = 0$ and $0.3$.}
	\label{fig_moment_Xi_Omega}
\end{figure}

\begin{figure}[ht!] 
 \includegraphics[width= 17 cm, height=18 cm]{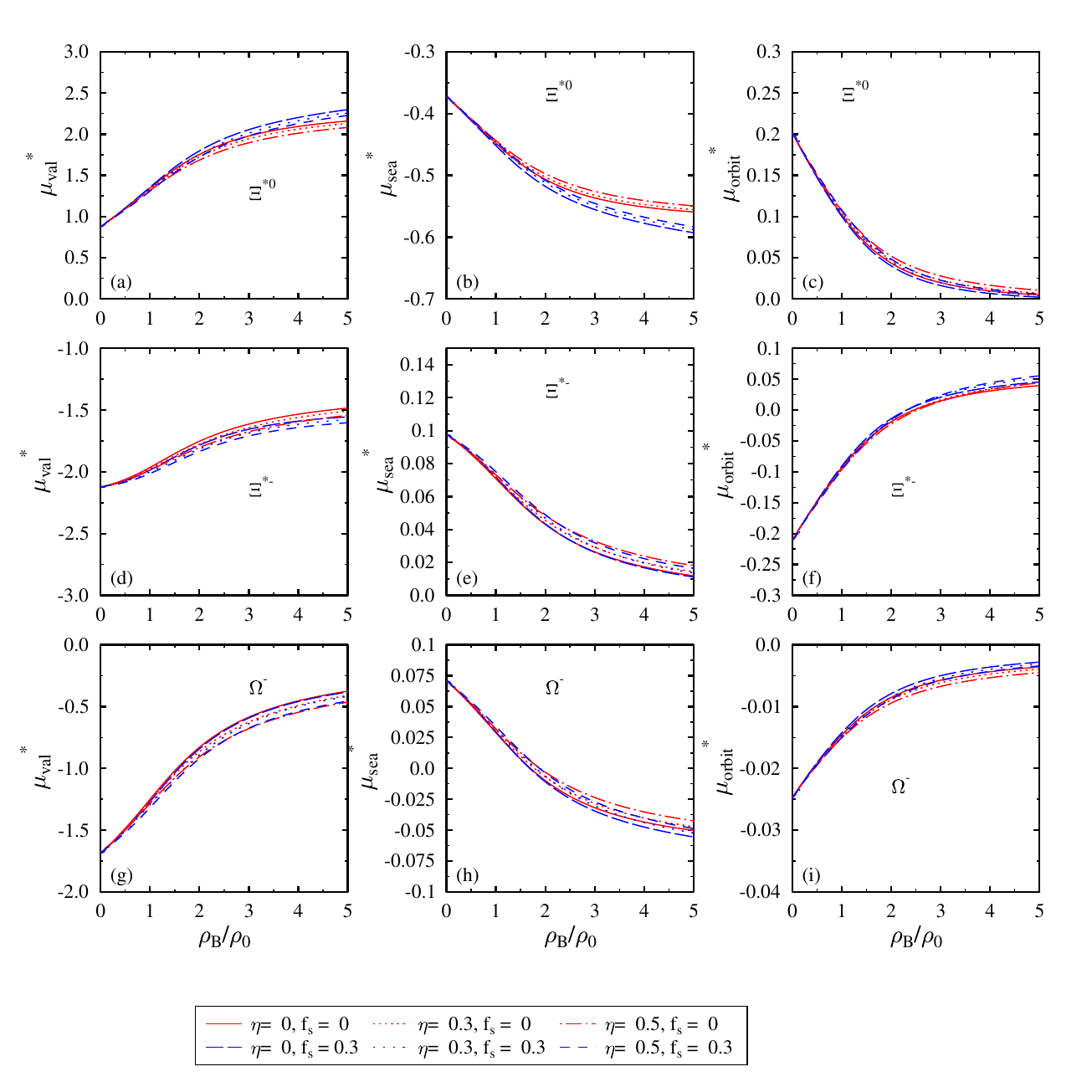}\hfill
	\caption{Contribution to total magnetic moment of
	$\Xi^{*}$ and $\Omega^-$ resonance baryons from   valance
	quarks, sea quarks and orbital moment of sea quark magnetic moments ($\mu_{val}^{*}$,$\mu_{sea}^{*}$ and $\mu_{orbit}^{*}$, respectively)   
	is shown as a function of baryon density $\rho_B$ (in units of nuclear saturation density $\rho_0$) at temperature $T = 0$ . Results are shown for isospin asymmetry $\eta = 0, 0.3, 0.5$ and strangeness fractions $f_s = 0$ and $0.3$.}
	\label{fig_moment_Xi_Omega_indi}
\end{figure}

The magnetic moments
of $\Sigma^{*+}$,  $\Sigma^{*0}$  and
 $\Sigma^{*-}$ as a function of $\rho_B$
 are plotted in \cref{fig_moment_Sigma}, whereas the individual 
 contributions due to valence, sea and orbital moment of sea quarks
 to the magnetic moments of
 $\Sigma^{*}$ is plotted
 in \cref{fig_moment_Sigma_indi}. Respective figures for decuplet members $\Xi^{*}$ $(\Xi^{*0}$ and $\Xi^{*-})$ and $\Omega^-$ baryons are \cref{fig_moment_Xi_Omega} and \cref{fig_moment_Xi_Omega_indi}.
The magnetic moments $\mu_{\Sigma^{*+}}^{*}$, $\mu_{\Sigma^{*0}}^{*}$
and $\mu_{\Sigma^{*-}}^{*}$
increase with baryon density $\rho_B$.
Understanding through the individual contributions,
as can be seen from \cref{fig_moment_Sigma_indi}, for
$\Sigma^{*+}$
 the magnetic moment of valence quarks $\mu_{val}^{*}$ increase with $\rho_B$ and dominate over the decrease of $\mu_{sea}^{*}$
 and $\mu_{orbit}^{*}$.
 In case of $\Sigma^{0*}$, 
 $\mu_{val}^{*}$ and
 $\mu_{orbit}^{*}$ increase whereas
 $\mu_{sea}^{*}$ 
 decreases with $\rho_B$ and the sum of former contribution dominates over latter.
Increase of strangeness fraction from $f_s = 0$ to $0.3$ enhances the magnitude of increase whereas  increasing the isospin asymmetry shows an opposite trend.
In case of $\Sigma^{-*}$ baryon,  the magnetic moment  $\mu_{orbit}^{*}$ increases with $\rho_B$ and dominates over the decrease of $\mu_{val}^{*}$
and $\mu_{sea}^{*}$. 


The magnetic moments of the decuplet baryons $\Xi^{*0}, \Xi^{*-}$ and $\Omega^{-}$ increase with increase in the baryon density and the increase is steeper at lower baryon densities. In the case of $\Xi^{*0}$, the magnetic moment of valence quark increases with density, whereas
sea quark and orbital moment of sea quark cause a small decrease as a function of $\rho_B$.
Increase in the strangeness fraction (keeping other parameters fixed) enhances the magnetic moment of $\Xi^{*0}$ and shows a decrease for
$\Xi^{*-}$. On the otherside,
increasing $\eta$ from 0 to 0.5 causes a decrease in the values of magnetic moment for both members of the $\Xi^{*}$ baryon multiplet.
At $\eta = 0$, values of $\mu_{\Omega^-}^{*}$ have a small difference  in nuclear and strange medium. However, increase of $\eta$ from 0 to 0.5 increases the difference in the values of magnetic moment of $\Omega^{-}$ baryons in the non-strange and strange medium.

\section{Summary}
	\label{sec:summary}
To summarize, in the present work we investigated the modification in the masses and magnetic moments of decuplet baryons in isospin asymmetric strange hadronic medium.
The in-medium effects were stimulated using the chiral SU(3) quark mean field model and the magnetic moments were calculated using the constituent chiral quark model. Isospin asymmetric effects are observed to be more significant for the baryons composed of light  $u$ or $d$ quarks, whereas 
strangeness effects became appreciable
for the baryons having strange quarks.  Increase of strangeness fraction in the medium enhances the drop in the masses of decuplet baryons whereas finite isospin asymmetry  tends to reduce this.
The behaviour of magnetic moments of decuplet baryons as a function of density of nuclear matter is different for different baryons and is determined by the individual contributions of valence quarks, sea quarks and orbital magnetic moment of sea quarks.
Both, finite isospin asymmetry and finite strangeness fraction of the medium are found to impact the in-medium values of magnetic moments of decuplet baryons.
The results of present work will be useful for different experimental facilities where dense matter with finite asymmetry is expected to be produced and also for those where internal structure of hadrons is  aimed to be explored.

	\begin{table}
	\begin{tabular}{|c|c|c|c|c|} 
	\hline 
	$k_0$ & $k_1$ & $k_2$ & $k_3$ & $k_4$  \\ 
	\hline 
	4.94 & 2.12 & -10.16 & -5.38 & -0.06  \\ 
	\hline
	\hline
	\hline 
	$\sigma_0$ (MeV) & $\zeta_0$ (MeV) & $\chi_0$ (MeV) & $\xi$ & $\rho_0$ ($\text{fm}^{-3}$)  \\ 
	\hline 
	-92.8 & -96.5 & 254.6 & 6/33 & 0.16  \\ 
	\hline
	\hline
	\hline 
	$g_{\sigma}^u=g_{\sigma}^d$ & $g_{\sigma}^s = g_{\zeta}^u=g_{\zeta}^d$ & $g_{\delta}^u$ & $g_{\zeta}^s = g_s$ & $g_4$\\ 
	\hline 
2.72 & 0 & 2.72 & 3.847
	 & 37.4 \\ 
	\hline
	\hline 
		$g_{\delta}^p = g_{\delta}^u$  & $g_{\omega}^N = 3g_{\omega }^u$  & $g_{\rho}^p$  & $m_{\pi}$ (MeV) & $m_K$ (MeV) \\
		\hline
		\hline 
			2.72 & 9.69 & 8.886 & 139 & 494  \\ 
			\hline
			\hline 
	\end{tabular}
	\caption{In above table values of various parameters used in the present work are given \cite{wang_nuc2001}.} \label{table1_para}
	\end{table}
	
\begin{sidewaystable}
\begin{tabular}{|c|c|c|c|c|c|c|c|c|c|c|c|c|c|}
\hline & \multirow{2}{*}{\begin{tabular}{l} 
Exp. values \cite{ Linde1998,Hagiwara2002}  \\
$\mu_B$ (In free space)
\end{tabular}} & \multicolumn{4}{|l|}{$\rho_B=0$} & \multicolumn{4}{|c|}{$\rho_B=\rho_0$} & \multicolumn{4}{|c|}{$\rho_B=3 \rho_0$} \\
\hline & & $\mu_{B,\mathrm{val}}^{*}$ & $\mu_{B,\text {sea}}^{*}$ & $\mu_{B,\text {orbital }}^{*}$ & $\mu_B^*$ & $\mu_{B,\mathrm{val}}^{*}$ & $\mu_{B, \text {sea }}^{*}$ & $\mu_{B,\text {orbital }}^{*}$ & $\mu_B^*$ & $\mu_{B,\mathrm{val}}^{*}$ & $\mu_{B,\text {sea }}^{*}$ & $\mu_{B,\text {orbital }}^{*}$ & $\mu_B^*$ \\
\hline$\mu_{\Delta^{++}}^*\left(\mu_N\right)$ & $ 4.5-7.5$ &
6.0 & -1.26 & 0.653 & 5.393 &
7.27 & -1.541 & 0.335 & 6.063 &
8.836 & -1.897 & 0.07 & 7.01 \\
\hline$\mu_{\Delta^{+}}^*\left(\mu_N \right)$ & - &
3.0 & -0.789 & 0.242 & 2.452 &
3.635 & -0.972 & 0.14 & 2.803 &
4.418 & -1.204 & 0.066 & 3.28 \\
\hline$\mu_{\Delta^0}^*\left(\mu_N \right)$ & - &
0.0 & -0.319 & -0.17 & -0.489 &
0.0 & -0.402 & -0.055 & -0.457 &
0.0 & -0.512 & 0.061 & -0.451 \\
\hline$\mu_{\Delta^{-}}^*\left(\mu_N \right)$ & - &
-3.0 & 0.151 & -0.581 & -3.43 &
-3.635 & 0.168 & -0.25 & -3.717 &
-4.418 & 0.181 & 0.056 & -4.181 \\
\hline$\mu_{\Sigma^{*+}}^*\left(\mu_N \right)$ & - &
3.439 & -0.816 & 0.427 & 3.05 &
4.148 & -0.964 & 0.218 & 3.402 &
4.993 & -1.133 & 0.045 & 3.905 \\
\hline$\mu_{\Sigma^{*-}}^*\left(\mu_N \right)$ & - &
-2.561 & 0.124 & -0.396 & -2.832 &
-2.746 & 0.117 & -0.171 & -2.801 &
-2.837 & 0.095 & 0.036 & -2.707 \\
\hline$\mu_{\Sigma^{* 0}}^*\left(\mu_N \right)$ & - &
0.439 & -0.346 & 0.016 & 0.109 &
0.701 & -0.424 & 0.023 & 0.301 &
1.078 & -0.519 & 0.04 & 0.599 \\
\hline$\mu_{\Xi^{* 0}}^*\left(\mu_N\right)$ & - &
0.878 & -0.373 & 0.201 & 0.706 &
1.346 & -0.448 & 0.102 & 1.0 &
1.978 & -0.537 & 0.02 & 1.461 \\
\hline$\mu_{\Xi^{*-}}^*\left(\mu_N\right)$ & - &
-2.122 & 0.098 & -0.21 & -2.235 &
-1.964 & 0.071 & -0.093 & -1.986 &
-1.613 & 0.026 & 0.015 & -1.572 \\
\hline$\mu_{\Omega^{-}}^*\left(\mu_N\right)$ & $-2.02 \pm 0.005$ &
-1.683 & 0.071 & -0.025 & -1.637 &
-1.251 & 0.029 & -0.015 & -1.237 &
-0.585 & -0.032 & -0.006 & -0.623 \\
\hline
\end{tabular}
\caption{Values of in-medium magnetic moments of decuplet baryons are tabulated above in symmetric nuclear matter ($\eta =0$ and $f_s =0$) at temperature $T=0$
and compared with values at $\rho_B =0$.}
\label{table:mub_eta0fs0}
\end{sidewaystable}

\begin{sidewaystable}
\begin{tabular}{|c|c|c|c|c|c|c|c|c|c|c|c|c|c|}
\hline & \multirow{2}{*}{\begin{tabular}{l} Exp. values \cite{ Linde1998,Hagiwara2002}  \\
$\mu_B$ (In free space)
\end{tabular}} & \multicolumn{4}{|l|}{$\rho_B=0$} & \multicolumn{4}{|c|}{$\rho_B=\rho_0$} & \multicolumn{4}{|c|}{$\rho_B=3 \rho_0$} \\
\hline & & $\mu_{B,\mathrm{val}}^{*}$ & $\mu_{B,\text {sea}}^{*}$ & $\mu_{B,\text {orbital }}^{*}$ & $\mu_B^*$ & $\mu_{B,\mathrm{val}}^{*}$ & $\mu_{B, \text {sea }}^{*}$ & $\mu_{B,\text {orbital }}^{*}$ & $\mu_B^*$ & $\mu_{B,\mathrm{val}}^{*}$ & $\mu_{B,\text {sea }}^{*}$ & $\mu_{B,\text {orbital }}^{*}$ & $\mu_B^*$ \\
\hline$\mu_{\Delta^{++}}^*\left(\mu_N\right)$ & $ 4.5-7.5$ &
6.0 & -1.26 & 0.653 & 5.393 &
7.192 & -1.524 & 0.351 & 6.019 &
8.642 & -1.852 & 0.097 & 6.886 \\
\hline$\mu_{\Delta^{+}}^*\left(\mu_N \right)$ & - &
3.0 & -0.789 & 0.242 & 2.452 &
3.608 & -0.963 & 0.148 & 2.793 &
4.347 & -1.182 & 0.083 & 3.248 \\
\hline$\mu_{\Delta^0}^*\left(\mu_N \right)$ & - &
0.0 & -0.319 & -0.17 & -0.489 &
0.0 & -0.399 & -0.055 & -0.454 &
0.0 & -0.503 & 0.07 & -0.434 \\
\hline$\mu_{\Delta^{-}}^*\left(\mu_N \right)$ & - &
-3.0 & 0.151 & -0.581 & -3.43 &
-3.633 & 0.169 & -0.258 & -3.722 &
-4.399 & 0.183 & 0.056 & -4.16 \\
\hline$\mu_{\Sigma^{*+}}^*\left(\mu_N \right)$ & - &
3.439 & -0.816 & 0.427 & 3.05 &
4.106 & -0.956 & 0.229 & 3.379 &
4.89 & -1.113 & 0.062 & 3.84 \\
\hline$\mu_{\Sigma^{*-}}^*\left(\mu_N \right)$ & - &
-2.561 & 0.124 & -0.396 & -2.832 &
-2.755 & 0.118 & -0.177 & -2.814 &
-2.863 & 0.099 & 0.035 & -2.728 \\
\hline$\mu_{\Sigma^{* 0}}^*\left(\mu_N \right)$ & - &
0.439 & -0.346 & 0.016 & 0.109 &
0.685 & -0.42 & 0.026 & 0.291 &
1.032 & -0.51 & 0.049 & 0.571 \\
\hline$\mu_{\Xi^{* 0}}^*\left(\mu_N\right)$ & - &
0.878 & -0.373 & 0.201 & 0.706 &
1.316 & -0.444 & 0.107 & 0.979 &
1.895 & -0.526 & 0.028 & 1.397 \\
\hline$\mu_{\Xi^{*-}}^*\left(\mu_N\right)$ & - &
-2.122 & 0.098 & -0.21 & -2.235 &
-1.986 & 0.073 & -0.096 & -2.009 &
-1.675 & 0.033 & 0.014 & -1.628 \\
\hline$\mu_{\Omega^{-}}^*\left(\mu_N\right)$ & $-2.02 \pm 0.005$ &
-1.683 & 0.071 & -0.025 & -1.637 &
-1.285 & 0.032 & -0.015 & -1.268 &
-0.678 & -0.024 & -0.007 & -0.709 \\
\hline
\end{tabular}
\caption{Values of in-medium magnetic moments of decuplet baryons are tabulated above in asymmetric nuclear matter ($\eta =0.5$ and $f_s =0$) at temperature $T=0$
and compared with values at $\rho_B =0$.}
\label{table:mub_eta5fs0}
\end{sidewaystable}

\begin{sidewaystable}
\begin{tabular}{|c|c|c|c|c|c|c|c|c|c|c|c|c|c|}
\hline & \multirow{2}{*}{\begin{tabular}{l} 
Exp. values \cite{ Linde1998,Hagiwara2002}  \\
$\mu_B$ (In free space)
\end{tabular}} & \multicolumn{4}{|l|}{$\rho_B=0$} & \multicolumn{4}{|c|}{$\rho_B=\rho_0$} & \multicolumn{4}{|c|}{$\rho_B=3 \rho_0$} \\
\hline & & $\mu_{B,\mathrm{val}}^{*}$ & $\mu_{B,\text {sea}}^{*}$ & $\mu_{B,\text {orbital }}^{*}$ & $\mu_B^*$ & $\mu_{B,\mathrm{val}}^{*}$ & $\mu_{B, \text {sea }}^{*}$ & $\mu_{B,\text {orbital }}^{*}$ & $\mu_B^*$ & $\mu_{B,\mathrm{val}}^{*}$ & $\mu_{B,\text {sea }}^{*}$ & $\mu_{B,\text {orbital }}^{*}$ & $\mu_B^*$\\
\hline$\mu_{\Delta^{++}}^*\left(\mu_N\right)$ & $ 4.5-7.5$ &
6.0 & -1.26 & 0.653 & 5.393 &
7.299 & -1.547 & 0.329 & 6.08 &
8.929 & -1.917 & 0.059 & 7.071 \\
\hline$\mu_{\Delta^{+}}^*\left(\mu_N \right)$ & - &
3.0 & -0.789 & 0.242 & 2.452 &
3.65 & -0.975 & 0.138 & 2.812 &
4.465 & -1.217 & 0.063 & 3.311 \\
\hline$\mu_{\Delta^0}^*\left(\mu_N \right)$ & - &
0.0 & -0.319 & -0.17 & -0.489 &
0.0 & -0.403 & -0.053 & -0.456 &
0.0 & -0.517 & 0.068 & -0.45 \\
\hline$\mu_{\Delta^{-}}^*\left(\mu_N \right)$ & - &
-3.0 & 0.151 & -0.581 & -3.43 &
-3.65 & 0.169 & -0.243 & -3.724 &
-4.465 & 0.182 & 0.072 & -4.21 \\
\hline$\mu_{\Sigma^{*+}}^*\left(\mu_N \right)$ & - &
3.439 & -0.816 & 0.427 & 3.05 &
4.175 & -0.97 & 0.214 & 3.418 &
5.12 & -1.161 & 0.037 & 3.996 \\
\hline$\mu_{\Sigma^{*-}}^*\left(\mu_N \right)$ & - &
-2.561 & 0.124 & -0.396 & -2.832 &
-2.766 & 0.118 & -0.167 & -2.816 &
-2.899 & 0.096 & 0.046 & -2.757 \\
\hline$\mu_{\Sigma^{* 0}}^*\left(\mu_N \right)$ & - &
0.439 & -0.346 & 0.016 & 0.109 &
0.704 & -0.426 & 0.024 & 0.301 &
1.11 & -0.532 & 0.042 & 0.62 \\
\hline$\mu_{\Xi^{* 0}}^*\left(\mu_N\right)$ & - &
0.878 & -0.373 & 0.201 & 0.706 &
1.356 & -0.452 & 0.1 & 1.004 &
2.055 & -0.555 & 0.016 & 1.516 \\
\hline$\mu_{\Xi^{*-}}^*\left(\mu_N\right)$ & - &
-2.122 & 0.098 & -0.21 & -2.235 &
-1.985 & 0.072 & -0.091 & -2.004 &
-1.655 & 0.026 & 0.021 & -1.608 \\
\hline$\mu_{\Omega^{-}}^*\left(\mu_N\right)$ & $-2.02 \pm 0.005$ &
-1.683 & 0.071 & -0.025 & -1.637 &
-1.269 & 0.03 & -0.014 & -1.254 &
-0.592 & -0.034 & -0.005 & -0.632 \\
\hline
\end{tabular}
\caption{Values of in-medium magnetic moments of decuplet baryons are tabulated above in symmetric strange matter ($\eta =0$ and $f_s =0.3$) at temperature $T=0$
and compared with values at $\rho_B =0$.}
\label{table:mub_eta0fs3}
\end{sidewaystable}
	
 \begin{sidewaystable}
 \begin{tabular}{|c|c|c|c|c|c|c|c|c|c|c|c|c|c|}
 \hline & \multirow{2}{*}{\begin{tabular}{l} 
 Exp. values \cite{ Linde1998,Hagiwara2002} \\
 $\mu_B$ (In free space)
 \end{tabular}} & \multicolumn{4}{|l|}{$\rho_B=0$} & \multicolumn{4}{|c|}{$\rho_B=\rho_0$} & \multicolumn{4}{|c|}{$\rho_B=3 \rho_0$} \\
\hline & & $\mu_{B,\mathrm{val}}^{*}$ & $\mu_{B,\text {sea}}^{*}$ & $\mu_{B,\text {orbital }}^{*}$ & $\mu_B^*$ & $\mu_{B,\mathrm{val}}^{*}$ & $\mu_{B, \text {sea }}^{*}$ & $\mu_{B,\text {orbital }}^{*}$ & $\mu_B^*$ & $\mu_{B,\mathrm{val}}^{*}$ & $\mu_{B,\text {sea }}^{*}$ & $\mu_{B,\text {orbital }}^{*}$ & $\mu_B^*$\\
\hline$\mu_{\Delta^{++}}^*\left(\mu_N\right)$ & $ 4.5-7.5$ &
6.0 & -1.26 & 0.653 & 5.393 &
7.299 & -1.547 & 0.329 & 6.08 &
8.929 & -1.917 & 0.059 & 7.071 \\
\hline$\mu_{\Delta^{+}}^*\left(\mu_N \right)$ & - &
3.0 & -0.789 & 0.242 & 2.452 &
3.65 & -0.975 & 0.138 & 2.812 &
4.465 & -1.217 & 0.063 & 3.311 \\
\hline$\mu_{\Delta^0}^*\left(\mu_N \right)$ & - &
0.0 & -0.319 & -0.17 & -0.489 &
0.0 & -0.403 & -0.053 & -0.456 &
0.0 & -0.517 & 0.068 & -0.45 \\
\hline$\mu_{\Delta^{-}}^*\left(\mu_N \right)$ & - &
-3.0 & 0.151 & -0.581 & -3.43 &
-3.65 & 0.169 & -0.243 & -3.724 &
-4.465 & 0.182 & 0.072 & -4.21 \\
\hline$\mu_{\Sigma^{*+}}^*\left(\mu_N \right)$ & - &
3.439 & -0.816 & 0.427 & 3.05 &
4.175 & -0.97 & 0.214 & 3.418 &
5.12 & -1.161 & 0.037 & 3.996 \\
\hline$\mu_{\Sigma^{*-}}^*\left(\mu_N \right)$ & - &
-2.561 & 0.124 & -0.396 & -2.832 &
-2.766 & 0.118 & -0.167 & -2.816 &
-2.899 & 0.096 & 0.046 & -2.757 \\
\hline$\mu_{\Sigma^{* 0}}^*\left(\mu_N \right)$ & - &
0.439 & -0.346 & 0.016 & 0.109 &
0.704 & -0.426 & 0.024 & 0.301 &
1.11 & -0.532 & 0.042 & 0.62 \\
\hline$\mu_{\Xi^{* 0}}^*\left(\mu_N\right)$ & - &
0.878 & -0.373 & 0.201 & 0.706 &
1.356 & -0.452 & 0.1 & 1.004 &
2.055 & -0.555 & 0.016 & 1.516 \\
\hline$\mu_{\Xi^{*-}}^*\left(\mu_N\right)$ & - &
-2.122 & 0.098 & -0.21 & -2.235 &
-1.985 & 0.072 & -0.091 & -2.004 &
-1.655 & 0.026 & 0.021 & -1.608 \\
\hline$\mu_{\Omega^{-}}^*\left(\mu_N\right)$ & $-2.02 \pm 0.005$ &
-1.683 & 0.071 & -0.025 & -1.637 &
-1.269 & 0.03 & -0.014 & -1.254 &
-0.592 & -0.034 & -0.005 & -0.632 \\
\hline
\end{tabular}
 \caption{Values of in-medium magnetic moments of decuplet baryons are tabulated above in symmetric strange matter ($\eta =0.5$ and $f_s =0.3$) at temperature $T=0$
 and compared with values at $\rho_B =0$.}
 \label{table:mub_eta5fs3}
 \end{sidewaystable}

\newpage
\vspace{1cm}
\hspace{5cm}{\Large{\textbf{Appendix}}}

The equations of motion for scalar fields $\sigma$, $\zeta$ the dilaton field, $\chi$, scalar iso-vector field, $\delta$, and, the vector fields $\omega$, $\rho$
and $\phi$ obtained by minimizing the  thermodynamic potential are written as (\cref{eq:therm_min1}) 
\begin{align}
\label{eq_sigma}
\frac{\partial \Omega}{\partial \sigma} = 
k_{0}\chi^{2}\sigma
-4k_{1}\left( \sigma^{2} \, + \, \zeta^{2} \, + \, \delta^2\right) \sigma \, - \,
2k_{2}\left(\sigma^{3} + 3\sigma\delta^2\right)\, - \, 2k_{3}\chi \sigma \zeta \, - \,
\frac{\xi }{3}\chi^{4}\left(\frac{2\sigma}{\sigma^2-\delta^2}\right)
 \nonumber \\
 \, + \, \frac{\chi^{2}}{\chi _{0}^{2}}m_{\pi }^{2}f_{\pi }
-\left( \frac{\chi }{\chi _{0}}\right)^{2}m_{\omega }\omega ^{2}
\frac{\partial m_{\omega }}{\partial \sigma }\, -\left( \frac{\chi }{\chi _{0}}\right)^{2}m_{\rho }\rho ^{2}
\frac{\partial m_{\rho }}{\partial \sigma }\, + \,
 \sum_{i} \frac{\partial M_{i}^{\ast }}
{\partial \sigma } \rho_i^s=0,~~~~~
\end{align}
\begin{align}
\label{zeta}
\frac{\partial \Omega}{\partial \zeta} & =
k_{0}\chi^{2}\zeta - 4k_{1}\left(\sigma^{2} \, + \,
\zeta ^{2} + \delta^2\right)
 \zeta \, - \, 4k_{2}\zeta ^{3} \, - \,
k_{3}\chi \sigma ^{2} \, 
- \, \frac{\xi\chi^{4}}{3\zeta } \, 
\nonumber \\&+ \,
\frac{\chi^{2}}{\chi _{0}^{2}} \left( \sqrt{2}m_{K}^{2}f_{K} \, - \,
\frac{1}{\sqrt{2}}m_{\pi }^{2}f_{\pi } \right)
+ \,
 \sum_{i=\Lambda, \Sigma^{\pm,0}, \Xi^{-,0}} \frac{\partial M_{i}^{\ast }}
{\partial \zeta } \rho_i^s=0,~~~~~ = 0,
\end{align}
\begin{align}
\label{scalar1}
\frac{\partial \Omega}{\partial \chi} =
 k_0\chi
\left(\sigma^2+\zeta^2+\delta^2\right)
-k_3\left(\sigma^2-\delta^2\right)\zeta+\frac{2\chi}{\chi _{0}^{2}}\left[  m_{\pi }^{2}f_{\pi }\sigma+\left( \sqrt{2}m_{K}^{2}f_{K} \, - \,
\frac{1}{\sqrt{2}}m_{\pi }^{2}f_{\pi } \right)\zeta\right] \nonumber \\
-\frac{\chi}{\chi_0^2} 
\left(m_\omega^2\omega^2+m_\rho^2\rho^2\right)
 +\left(4k_4+1+{\rm ln}\frac{\chi^4}{\chi_0^4} -
\frac{4\xi}
3{\rm ln}\left(\left(\frac{\left(\sigma^2-\delta^2\right)\zeta}{\sigma_0^2\zeta_0}\right)\right)\right)\chi^3 
 =0,
\end{align}
\begin{align}
\label{eq_delta}
\frac{\partial \Omega}{\partial \delta} & =
k_{0}\chi^{2}\delta
-4k_{1}\left( \sigma^{2} \, + \, \zeta^{2} \, + \, \delta^2\right) \delta \, - \,
2k_{2}\left(\delta^{3} + 3\sigma^2\delta\right)\, + \, 2k_{3}\chi \delta \zeta \, - \,
\frac{\xi }{3}\chi^{4}\left(\frac{2\delta}{\sigma^2-\delta^2}\right)
 \nonumber \\
 & - \,
 \sum_{i}
  g_{\delta i} \rho_i^s=0,~~~~~
\end{align}
\begin{align}
\frac{\partial \Omega}{\partial \omega} =
 \left(\frac{\chi^2}{\chi_0^2}\right) 
m_\omega^2\omega+4g_4
\omega^3+12g_4\omega\rho^2-\sum_{i} g_{\omega i} \rho_i=0, \label{vector1}
\end{align}

\begin{align}
 \frac{\partial \Omega}{\partial \rho} = \left(\frac{\chi^2}{\chi_0^2}\right) 
m_\rho^2\rho+4g_4
\rho^3+12g_4\omega^2\rho-\sum_{i} {g_{\rho i}} \rho_i=0, \label{vector2}
\end{align}
and
\begin{align}
\frac{\partial \Omega}{\partial \phi}= \left(
\frac{\chi^2}{\chi_0^2} \right) m_\phi^2\phi+8g_4\phi^3-
\sum_{i}g_{\phi\,i}\rho_{i}=0,
 \label{eqe_phi}  
\end{align}
respectively.
The number (vector) density, $\rho_{i}$, and scalar density, $\rho_{i}^{s}$, of baryons appearing in above equation
are given as 
\begin{align}
\rho_{i} = \gamma_{i} \int\frac{d^{3}k}{(2\pi)^{3}}  
\Big(f_i(k)-\bar{f}_i(k)
\Big),
\label{rhov0}
\end{align}
 and
\begin{eqnarray}
\rho_{i}^{s} = \gamma_{i} \int\frac{d^{3}k}{(2\pi)^{3}} 
\frac{m_{i}^{*}}{E^{\ast}_i(k)} \Big(f_i(k)+\bar{f}_i(k)
\Big),
\label{rhos0}
\end{eqnarray}
respectively, where $f_i(k)$ and $\bar{f}_i(k)$ represent the Fermi distribution functions at finite temperature for fermions and anti-fermions and are expressed as 
\bea
f_i(k) &=& \frac{1}{1+\exp\left[(E^*_{i}(k) 
-\nu^{*}_{i})/k_BT \right]}~~~ \text{and}~~~
\bar{f}_i(k) = \frac{1}{1+\exp\left[( E^*_{i}(k) 
+\nu^{*}_{i})/k_BT\right]}~.
\label{dfp}
\eea  

		\bibliographystyle{elsarticle-num}


\end{document}